\documentclass[%
 reprint,
superscriptaddress,onecolumn,
 amsmath,amssymb,
 aps,
floatfix
]{revtex4-2}

\usepackage[export]{adjustbox}
\usepackage{natbib}
\usepackage[scr=rsfs]{mathalpha}
\usepackage{graphicx}% Include figure files
\usepackage[T1]{fontenc}
\usepackage{outlines}
\usepackage{epstopdf}
\usepackage{epsfig}
\usepackage{xcolor}
\usepackage{subcaption}
\usepackage{dcolumn}% Align table columns on decimal point
\usepackage{bm}
\usepackage{bbold}
\DeclareMathOperator{\tr}{tr}
\usepackage{caption}
\makeatletter
\renewcommand\@makecaption[2]{%
  \par
  \vskip\abovecaptionskip
  \begingroup
   \small\rmfamily
    \begingroup
     \samepage
     \flushing
     \let\footnote\@footnotemark@gobble
     \@make@capt@title{#1}{#2}\par
    \endgroup
  \endgroup
  \vskip\belowcaptionskip
}
\makeatother

\begin{document}

\preprint{APS/123-QED}

\title{Impact of asymmetries in valences and diffusivities on the transport of a binary electrolyte in a charged cylindrical pore}% Force line breaks with \\

\author{Filipe Henrique}
\affiliation{Department of Chemical and Biological Engineering, University of Colorado, Boulder}
\author{Pawel J. Zuk}%
\affiliation{%
Institute of Physical Chemistry, Polish Academy of Sciences,  Warsaw, Poland}%
\affiliation{%
Department of Physics, Lancaster University, Lancaster, United Kingdom}%
\author{Ankur Gupta}
\email{Corresponding author:
ankur.gupta@colorado.edu}
\affiliation{Department of Chemical and Biological Engineering, University of Colorado, Boulder}

\date{\today}% It is always \today, today,


\begin{abstract}

Ion transport in porous media is present in a wealth of technologies, e.g.,  energy storage devices such as batteries and supercapacitors, and environmental technologies such as electrochemical carbon capture and capacitive deionization. Recent studies on flat plate electrodes have demonstrated that asymmetries in ions properties, such as valences and diffusivities, lead to intriguing and counter-intuitive physical phenomena. Yet, the consequences of such asymmetries to ion transport have seldom been explored in porous geometries. To bridge this knowledge gap, we employ direct numerical simulations to solve Poisson-Nernst-Planck equations inside a cylindrical pore for a binary electrolyte with arbitrary valences and diffusivities. Next, we conduct a perturbation expansion in the limit of small potential and derive equations for charge and salt transport under confinement. We obtain good agreement between the perturbation analysis and the direct numerical simulations. Our analysis reveals that the charge and the salt transport are coupled with each other. Further,  the coupling between the charge and salt transport processes is enhanced with an increase in valence and diffusivity asymmetries of ions. We observe that the mismatch of the ionic diffusivities induces a non-trivial salt dynamics, producing either transient depletion or enhancement of salt in the pore.  In the regime of high static diffusion layer conductance, we obtain an analytical solution to our perturbation model. The solution elucidates how electrolyte asymmetry induces two charging timescales that are set by the relative pore size. In the overlapping-double-layer regime, these timescales reduce to the diffusion times of each ion such that the transport of the two ions appears to be decoupled. Overall, our work underscores that the asymmetry in cation and anion diffusivities fundamentally alters the behavior of ionic transport inside a charged cylindrical pore and opens up new avenues of research on electrolyte transport in porous materials. 

\end{abstract}

\maketitle

%\tableofcontents

\section{Introduction}\label{sec:int}

Ion transport near charged solid surfaces is crucial in several applications, e.g., energy storage devices such as supercapacitors \cite{bazant2004diffuse,biesheuvel2010nonlinear,kondrat2012effect,kondrat2014accelerating,smith2016electrostatic} and hybrid capacitors \cite{simon2008materials,biesheuvel2011diffuse,biesheuvel2012electrochemistry,simon2020perspectives}, desalination and ion separation techniques such as capacitive deionization \cite{porada2012water,porada2013review,biesheuvel2010membrane} and Faradaic capacitive deionization \cite{zhang2018faradaic,he2018theory}, electrochemical carbon capture using ionic liquids \cite{zhang2012carbon}, as well as nanofluidic devices to enhance the detection sensitivity of low abundance protein immunoassays   \cite{ko2012nanofluidic}. An important characteristic of ion transport in these physical systems is the formation of charged regions -- called electrical double layers (EDLs) -- next to the solid surfaces. In these regions, both diffusive and electromigrative fluxes may play a role \cite{peters2016analysis,gupta2020charging,henrique2022charging} in electrolyte transport. Typically, the thickness of EDLs ranges from $0.5-10$ nm \cite{kornyshev2007double,huang2008theoretical}, though some studies have argued that it can be significantly larger for ionic liquids \cite{gebbie2013ionic,smith2015influence,gavish2018solvent,avni2019charge}. 
\par{} The most common physical setup for ion transport studied in the literature is the flat-plate geometry, where researchers generally motivate the analysis by assuming the thin-double-layer limit, i.e., that the plate separation is much larger than the EDL thickness, to evaluate the ion fluxes, electric  potential, and concentration profiles   \cite{bazant2004diffuse,kilic2007steric1,kilic2007steric2,feicht2016discharging}. A recent focus in the literature for the flat-plate geometry is the impact of valence and diffusivity asymmetries in ion transport  \cite{balu2018role,amrei2018oscillating,balaji2019ac,amrei2019asymmetric,amrei2020perturbation,balu2021thin,balu2022electrochemical}. We remark on two effects that occur due to a diffusivity contrast: (i) for an applied AC potential, a steady long-range electric field, also known as asymmetric rectified electric field (AREF), is observed for electrolytes with a diffusivity contrast \cite{amrei2018oscillating,amrei2019asymmetric,amrei2020perturbation}; (ii) a long-range repulsion force between oppositely charged surfaces upon application of an AC field \cite{richter2020ions}.  
\par{} Although the flat-plate geometry and  thin-double-layer assumptions have enabled fundamental investigations into various aspects of electrolyte transport, in many of the applications described above transport occurs within porous media, where both the geometry and length scales differ from the flat-plates setup. The porous structure is thus commonly approximated by a narrow cylindrical pore such that the length $\ell_p$ is much greater than its radius  $a_p$ \cite{biesheuvel2010nonlinear,biesheuvel2011diffuse,zhao2012time,alizadeh2017multiscale,alizadeh2017multiscale2,alizadeh2019impact,gupta2020charging,henrique2022charging,yang2022direct}. While the approximation of a porous electrode by a cylindrical pore is an improvement from the flat-plate geometry, many of these reports also assume the thin-double-layer limit, where the Debye length $\lambda_D$ follows $a_p/\lambda_D \gg 1$ \cite{biesheuvel2010nonlinear,biesheuvel2011diffuse,zhao2012time}. However, the characteristic pore width in many scenarios is comparable to the double-layer thickness. For instance, in supercapacitors, where energy is stored due to transport of ions to the pores within the electrodes, a typical pore radius is on the order of nanometers \cite{simon2008materials,simon2020perspectives}, which is comparable to the Debye length. Similarly, in capacitive desalination devices, pores with sizes on the order of $1$ nm, with comparable Debye length and pore radius, contribute to ion removal \cite{porada2013review,he2018theory}.
\par{} To approach experimentally relevant pore sizes, recent literature on cylindrical pores has attempted to go beyond the thin-double-layer limit. Specifically, Alizadeh and Mani proposed a theoretical framework to handle electro-osmotic flows for cylindrical pores for charged EDLs without imposing any restrictions on $a_p/\lambda_D$ \cite{alizadeh2017multiscale,alizadeh2017multiscale2,alizadeh2019impact}. Our recent work proposed a time-dependent solution for charging of EDLs, where we discovered pronounced changes in the potential and ion concentrations near the mouth of the pore when $a_p/\lambda_D \gtrsim O(1)$. We have further demonstrated the modification required for moderate to overlapping double layers on the effective transmission line circuit. Finally, we also found that the interplay of electromigrative and diffusive fluxes in overlapping double layers may cause a significant change in the charging timescale of pores \cite{gupta2020charging,henrique2022charging}.  Our results have been verified and recently extended to large potentials through direct numerical simulations by Yang et al. \cite{yang2022direct}. Other approaches utilized in the literature invoke classical density functional theory \cite{tomlin2021impedance} or molecular dynamics simulations \cite{kondrat2012effect,kondrat2014accelerating}.    
% the ef
\par{} To the best of our knowledge, the only approach that describes the transport in a cylindrical pore for arbitrary ionic diffusivities and valences utilizes the modified Donnan potential framework  \cite{biesheuvel2012electrochemistry}. While the modified Donnan potential concept is useful, it relies on a lumped-parameter approach and overlooks the spatial variations inside the porous structure. In our work, we directly solve the Poisson-Nernst-Planck (PNP) in the Debye-H{\"u}ckel limit of low applied potentials without imposing any restrictions on $a_p/\lambda_D$. In particular, to capture the effect of diffusivity and valence asymmetry, we extend our framework proposed for symmetric electrolytes \cite{gupta2020charging,henrique2022charging}, and derive effective transport equations for charged cylindrical pores. For simplicity, we assume the electrodes to be blocking, neglecting the possibility of heterogeneous chemical reactions in this contribution. 
\par{} Specifically, we perform an asymptotic expansion of the PNP equations for asymmetric electrolytes using the small applied potential as the asymptotic parameter. From our analysis, we find key features of asymmetry at low potentials: 
\begin{enumerate}
    \item Unlike symmetric electrolytes \cite{gupta2020charging,henrique2022charging}, the potential profile in the static diffusion layer is not necessarily linear;
    \item The unequal rates of transport of cations and anions produce salt (i.e., the  sum of cation and anion concentrations) changes in the pore even at low potentials.
\end{enumerate}  
These unequal rates also set two characteristic timescales of the process which depend on relative pore size, which we denote by $\kappa=a_p/\lambda_D$. Furthermore, even though the steady-state charge is the same for all valences and diffusivities, the overall timescale required to charge the pore can be controlled by these transport coefficients, i.e., the charging timescale is no longer solely determined by the capacitance of the pore.

\section{Problem Formulation}\label{sec:form}

\subsection{Governing Equations}

\begin{figure}[h]
    \centering
    \includegraphics[max width=0.5\textwidth]{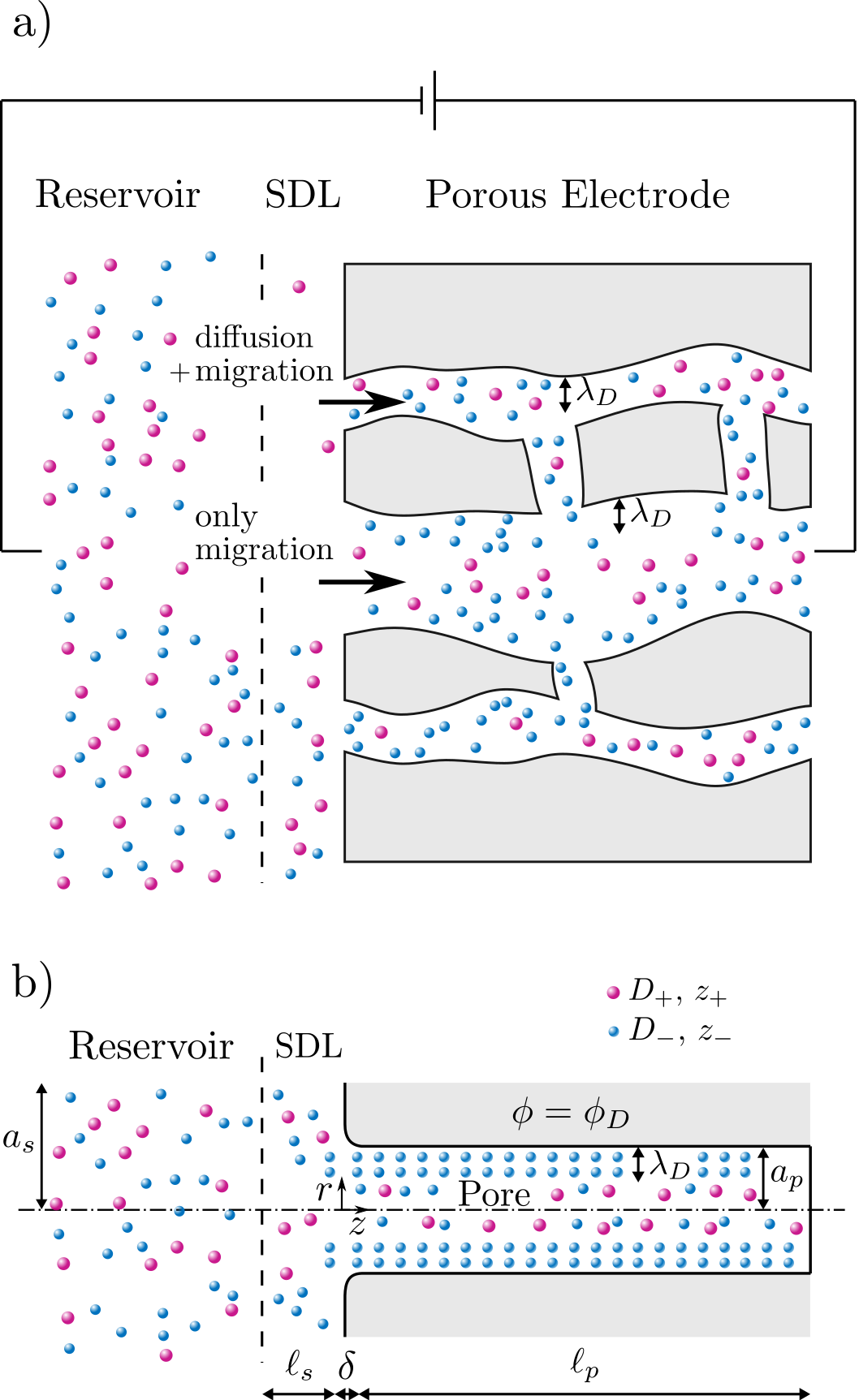}
    \caption{\textbf{Schematic of the problem}. a) Schematic illustration of ion transport in an asymmetric electrolyte with ionic valences $z_\pm$ and diffusivities $D_\pm$ in a porous electrode. b) Simplified representation of the electrode by a cylindrical pore under an applied potential $\phi_D$. The pore region of radius $a_p$ and length $\ell_p$ is connected to a cylindrical electroneutral static diffusion layer (SDL) of radius $a_s$ and length $
    \ell_s$. The length $\delta$ of the transition region follows $\delta/\ell_p \ll 1$. The Debye length $\lambda_D$ may be comparable to the radius of the pore.}
    \label{fig:pore}
\end{figure}

We seek to elucidate the effects of unequal ion diffusivities $D_\pm$ and valences $z_\pm$ ($z_+>0$ and $z_-<0$) on the charging of a porous electrode with arbitrary relative pore size $\kappa$. As illustrated in Fig. \ref{fig:pore}a, the electrode may consist of tortuous connected pores. Importantly, in pores of different sizes, distinct ion transport mechanisms may be at play. In large pores, with relative pore sizes $\kappa\gg 1$, transport is dominated by electromigration. However, for small pores, with relative pore sizes $\kappa\sim 1$, transport is set by a balance of diffusion and  electromigration. In the interest of investigating the effects of pore size in the presence of ionic valence and diffusivity asymmetries, we utilize a simplified representation of electrode pores as individual non-interacting cylindrical pores, shown in Fig. \ref{fig:pore}b. We assume that the length of the pores is much greater than their radius, $\ell_p\gg a_p$, as is characteristic of experimental setups \cite{pell2000analysis,zhang2014highly}. We solve for electrolyte transport when an electrical potential $\phi_D$ is applied to the surface of the pore. We note that the application of the potential on the surface initiates a transport of oppositely-charged ions into the pore, and of similarly charged ions out of the pore. We contrast this scenario with the analysis of Alizadeh and Mani \cite{alizadeh2017multiscale, alizadeh2019impact}, who focused on the scenario of a symmetric electrolyte where the pores are already charged. That is the reason why they do not consider a time-dependent analysis of the electric double layers. Fig. \ref{fig:pore}b shows our division of the electrolyte domain into four regions:
\begin{enumerate}
    \item an electroneutral reservoir ($z<-\ell_s$ where $\ell_s$ is the length of the static diffusion layer (SDL)), which is unaffected by the applied potential, maintaining a reference potential and the concentration of ions resulting from dissociation;
    \item the static diffusion layer (SDL), cylindrical with radius $a_s$ (defined by $(r,z)\in [0,a_s]\times [-\ell_s,0)$), which is also approximately electroneutral as we will discuss in Sec. \ref{sec:sdl}, but where an electric field develops in response to the applied potential;
    \item a transition region that connects a charged pore to the SDL, and which presents significant changes in charge density and electric potential for finite-sized double layers \cite{gupta2020charging, henrique2022charging}. We assume $\delta/\ell_p \ll 1$ under the framework of long pores \cite{henrique2022charging};
    \item the pore region itself, $(r,z)\in [0,a_p]\times [0,\ell_p]$ which charges radially and axially in response to the applied potential $\phi_D$;
\end{enumerate} 

In this setting, the bulk has a reference potential $\phi=0$ and the ion concentrations in the bulk satisfy electroneutrality, i.e., $c_{\pm}=\mp z_{\mp}c_{\infty}$, where $c_\infty$ is the concentration of the electrolyte before dissociation. For instance, CaCl$_2$ with $z_+=2$ and $z_-=-1$ will follow $c_+ = c_{\infty}$ and $c_-=2 c_{\infty}$. The PNP equations in the pore for such an electrolyte are \cite{deen2012analysis, newman2012electrochemical}
\begin{equation}
    \begin{cases}
    \dfrac{\partial c_\pm}{\partial t}=D_\pm \nabla^2 c_\pm+\dfrac{z_\pm D_\pm e}{k_BT}\nabla\cdot\left(c_\pm\nabla\phi\right),\\[10pt]
    -\varepsilon\nabla^2\phi=e(z_+c_++z_-c_-),
    \end{cases}
\end{equation}
where $t$ is time, $e$ is the charge of an electron, $k_B$ is the Boltzmann constant, $T$ is temperature and $\varepsilon$ is the electrical permittivity of the electrolyte, assumed here to be constant. Similarly to Ref. \cite{amrei2020perturbation}, it is useful for us to define the following parameters: arithmetic mean diffusivity in the pore\footnote{A subtle difference being that Ref. \cite{amrei2020perturbation} defines a harmonic mean diffusivity $D_p$, whereas in our work it is more convenient to define an arithmetic mean diffusivity $D_p$.}
\begin{equation}
    D_p=\dfrac{D_++D_-}{2}
\end{equation} 
and a coefficient of diffusivity contrast,
\begin{equation}
    \beta=\dfrac{D_+-D_-}{D_++D_-},
\end{equation}
furnishing the relation
\begin{equation}
    \dfrac{D_\pm}{D_p}=1\pm\beta.
\end{equation}
We note that the values of the ionic diffusivities in the SDL might differ from those in the pore, where confinement effects might play a role \cite{rawlings2002chemical,henrique2022charging}. We assume that both diffusivities change by the same ratio going from pore to SDL, such that in the latter they are multiplied by the factor $D_s/D_p$, where $D_s$ is the arithmetic mean of the diffusivities in the SDL. We also define an average valence
\begin{equation}
    z_\mathrm{avg}=\dfrac{z_+-z_-}{2}
\end{equation} 
and a contrast of valences
\begin{equation}
    \gamma=\dfrac{z_++z_-}{z_+-z_-},
\end{equation}
yielding
\begin{equation}
    \dfrac{z_\pm}{z_\mathrm{avg}}=\gamma\pm 1,
\end{equation}
and a reference concentration
\begin{equation}
    c_0=z_+z_-(z_--z_+)c_\infty/2z_\mathrm{avg}^2,
    \label{eq:c0}
\end{equation}
chosen such that the Debye length takes the form $\lambda_D=[\varepsilon k_BT/(2z_\mathrm{avg}^2e^2c_0)]^{1/2}$; see Ref. \cite{amrei2020perturbation}. We nondimensionalize time as $\tau=tD_p/\ell_p^2$ based on the pore arithmetic mean ionic diffusivity, the axial coordinate as $Z=z/\ell_p$, and the  radial coordinate as $R=r/a_p$, such that $\tilde{\nabla}=\ell_p\nabla=\hat{\boldsymbol{e}}_Z\partial/\partial Z+(\ell_p/a_p)\hat{\boldsymbol{e}}_R\partial/\partial R$ for the gradient operator, the potential as $\Psi=z_\mathrm{avg}e\phi/(k_BT)$, and concentration as $\tilde{c}=c/c_0$. Using these scales and defining dimensionless charge and salt densities respectively by $\rho=e(z_+c_++z_-c_-)/(2ez_\mathrm{avg}c_0)$ and $s=(c_++c_-)/(2c_0)$, the dimensionless PNP equations take the form
%\begin{widetext}
\begin{equation}
\begin{cases}
    \dfrac{\partial\rho}{\partial\tau}=-\tilde{\nabla}\cdot\mathbf{J},\\[10pt]
    \dfrac{\partial s}{\partial\tau}=-\tilde{\nabla}\cdot\mathbf{W},
    \end{cases}
    \label{Eq: pnp}
\end{equation}
%\end{widetext}
where the dimensionless charge and salt fluxes read
\begin{equation}
\begin{cases}
    \mathbf{J}=-\tilde{D}\left\{(1+\beta\gamma)\tilde{\nabla}\rho+\beta(1-\gamma^2)\tilde{\nabla}s+[\beta(1+\gamma^2)+2\gamma]\rho\tilde{\nabla}\Psi+(1+\beta\gamma)(1-\gamma^2)s\tilde{\nabla}\Psi\right\},\\[10pt]
    \mathbf{W}=-\tilde{D}\left[\beta\tilde{\nabla}\rho+(1-\beta\gamma)\tilde{\nabla}s+(1+\beta\gamma)\rho\tilde{\nabla}\Psi+\beta(1-\gamma^2)s\tilde{\nabla}\Psi\right],
\end{cases}
\label{Eq: fluxes}
\end{equation}
with $\tilde{D}=1$ in the pore and $\tilde{D}=D_s/D_p$ in the SDL. 

In Sec. \ref{sec:num}, we will perform Direct Numerical Simulations (DNS) of the full equations \eqref{Eq: pnp} and \eqref{Eq: fluxes} with initial conditions
\begin{align}
&\rho(R,Z,\tau=0)=0,\\
&s(R,Z,\tau=0)=1/(1-\gamma^2),\\
&\Psi(R,Z,\tau=0)=\begin{cases}
\Psi_D(1+Z\ell_p/\ell_s),\quad Z\in(-\ell_s/\ell_p,0),\\
\Psi_D,\quad Z\in [0,1].
\end{cases}
\end{align}
We utilize boundary conditions representing the reservoir 
\begin{align}
&\rho(R,Z=-\ell_s/\ell_p,\tau)=0,\\
&s(R,Z=-\ell_s/\ell_p,\tau)=1/(1-\gamma^2),\\
&\Psi(R,Z=-\ell_s/\ell_p,\tau)=0,
\end{align}
the blocking surfaces of the porous electrode,
\begin{equation}
    \dfrac{\partial \rho}{\partial Z}\bigg|_{Z=1}=\dfrac{\partial s}{\partial Z}\bigg|_{Z=1}=\dfrac{\partial \Psi}{\partial Z}\bigg|_{Z=1}=0,\quad R\in [0,1]
\end{equation}
and
\begin{equation}
    J_R(R=1,Z,\tau)=W_R(R=1,Z,\tau)=0,\quad Z\in [0,1],
\end{equation}
and require the absence radial fluxes in or out of the SDL for non-interacting pores \cite{henrique2022charging},
\begin{equation}
    \dfrac{\partial \rho}{\partial R}\bigg|_{R=a_s/a_p}=\dfrac{\partial s}{\partial R}\bigg|_{R=a_s/a_p}=\dfrac{\partial \Psi}{\partial R}\bigg|_{R=a_s/a_p}=0,\quad Z\in (-\ell_s/\ell_p,0).
    \label{eq:fluxBCsdl}
\end{equation}
The applied potential at the surface of the pore gives
\begin{equation}
    \Psi(R=1,Z,\tau)=\Psi_D,\quad Z\in [0,1].
    \label{eq:BCpsi}
\end{equation}
Eqs. \eqref{Eq: pnp}--\eqref{eq:BCpsi} are solved by the finite-volume method using OpenFOAM \cite{weller1998tensorial,jasak2007openfoam}. More details of the procedure and geometry are found in the Supplemental Material of Ref. \cite{gupta2020charging}. These simulations present a high computational cost, therefore in the remainder of the paper, we pursue a reduced-order asymptotic model that captures the dynamics of pore charging that reveals the effects of asymmetries in valences and diffusivities at low computational cost, and use the DNS results for validation of the model.

\subsection{Perturbation Analysis}

In the interest of obtaining expressions amenable to analytical methods, we are particularly interested in the linear response of the system at low applied potentials, $|\Psi_D|\ll 1$. In this limit, we perform regular perturbation expansions of the dependent variables,
\begin{equation}
\begin{cases}
\rho=\rho_1\Psi_D+O(\Psi_D^2),\\[5pt] \Psi=\Psi_1\Psi_D+O(\Psi_D^2),\\[5pt] s=\dfrac{1}{1-\gamma^2}+s_1\Psi_D+O(\Psi_D^2),    
\end{cases}
\end{equation}
where the zeroth-order terms follow from the response in the absence of an applied electric field, when they are equal to the bulk properties. Substituting this expansion, we have
\begin{equation}
\begin{cases}
    \dfrac{\partial\rho_1}{\partial\tau}=\tilde{D}\left[(1+\beta\gamma)(\tilde{\nabla}^2\rho_1+\tilde{\nabla}^2\Psi_1)+\beta(1-\gamma^2)\tilde{\nabla}^2s_1\right],\\[10pt]
    \dfrac{\partial s_1}{\partial\tau}=\tilde{D}\left[\beta(\tilde{\nabla}^2\rho_1+\tilde{\nabla}^2\Psi_1)+(1-\beta\gamma)\tilde{\nabla}^2s_1\right],\\[5pt]
    -\tilde{\nabla}^2\Psi_1=\left(\dfrac{\ell_p}{\lambda_D}\right)^2\rho_1.
    \end{cases}
    \label{eq:PNPpert}
\end{equation}
Next, we write the appropriate simplifications of the PNP equations in each regions described earlier.

\subsubsection{Static Diffusion Layer}\label{sec:sdl}

In this non-interacting pore model, we neglect radial fluxes at the radial boundaries of the SDL, i.e., $\partial\Psi_1/\partial R|_{R=a_s/a_p}=0$, see Eq. (\ref{eq:fluxBCsdl}). Coupled with the symmetry of the problem with respect to an inversion of the $R$ axis, yielding $\partial\Psi_1/\partial R|_{R=0}=0$, we can assume the independence of the potential profile on the radial coordinate, simplifying Poisson's equation in (\ref{eq:PNPpert}) to
\begin{equation}
    -\left(\dfrac{\lambda_D}{\ell_p}\right)^2\dfrac{\partial^2\Psi_1}{\partial Z^2}=\rho_1.
    \label{}
\end{equation}
Furthermore, with $\lambda_D=O(a_p)$ at most for overlapping double layers, the long-pore assumption allows us to neglect the left-hand side of Poisson's equation, implying electroneutrality,
\begin{equation}
    \rho_1=0.
\end{equation}
The PNP equations \eqref{eq:PNPpert} in the SDL with no radial dependence and electroneutrality simplify to
\begin{equation}
\begin{cases}
    (1+\beta\gamma)\dfrac{\partial^2\Psi_1}{\partial Z^2}+\beta(1-\gamma^2)\dfrac{\partial^2s_1}{\partial Z^2}=0,\\[10pt]
    \dfrac{\partial s_1}{\partial\tau}=\dfrac{D_s(1-\beta^2)}{D_p(1+\beta\gamma)}\dfrac{\partial^2s_1}{\partial Z^2},
\end{cases}
\label{eq:pnpsdl}
\end{equation}
where $D_s(1-\beta^2)/[D_p(1+\beta\gamma)]$ is the dimensionless ambipolar diffusivity in the SDL. Notably, though salt inhomogeneity affects the potential profile, the salt and potential equations may be uncoupled in this region. We remark that, in contrast to our previous work on pore-size effects for symmetric electrolytes \cite{henrique2022charging}, in the asymmetric case the second derivative of potential is not negligible, so we cannot assume a linear potential profile in the SDL. Consequently, there is also a dynamics of salt migration in this region. Nonlinearity of the potential profile in the SDL has also been recently reported for symmetric electrolytes at large potentials by Yang  et. al. \cite{yang2022direct}.

In the next section, we will use pursue averaged equations for axial ion transport in the pore. In this context, it should be noted that Eq. \eqref{eq:pnpsdl} can be trivially averaged over its cross-section in a similar fashion to yield
\begin{equation}
    \begin{cases}
    (1+\beta\gamma)\dfrac{\partial^2\bar{\Psi}_1}{\partial Z^2}+\beta(1-\gamma^2)\dfrac{\partial^2\bar{s}_1}{\partial Z^2}=0,\\[10pt]
    \dfrac{\partial \bar{s}_1}{\partial\tau}=\dfrac{D_s(1-\beta^2)}{D_p(1+\beta\gamma)}\dfrac{\partial^2\bar{s}_1}{\partial Z^2}
    \end{cases}
    \label{eq:pnpsdlav}
\end{equation}
in the SDL, bars representing cross-sectional averages of the variables, as further discussed in the following section.

\subsubsection{Pore Region}

The applied potential at the pore surface imposes a radial dependence of the properties inside the pore. Under the assumption of long pores, $\ell_p\gg a_p$, we require radial equilibrium to hold within the pore, i.e., $\partial (R J_{1R})/\partial R=\partial (R W_{1R})/\partial R=0$.  Integrating these equations in $R$ in the absence of surface reactions yields $J_{1R}=W_{1R}=0$, i.e.,
\begin{equation}
    \begin{cases}
    (1+\beta\gamma)\left(\dfrac{\partial\rho_1}{\partial R}+\dfrac{\partial\Psi_1}{\partial R}\right)+\beta(1-\gamma^2)\dfrac{\partial s_1}{\partial R}=0,\\[10pt]
    \beta\left(\dfrac{\partial\rho_1}{\partial R}+\dfrac{\partial\Psi_1}{\partial R}\right)+(1-\beta\gamma)\dfrac{\partial s_1}{\partial R}=0.
    \end{cases}
    \label{eq:rad_equil}
\end{equation}
Solving the system of Eqs. \eqref{eq:rad_equil} and integrating from the centerline, denoted by the index $m$, to an arbitrary radial position in the pore, we establish the relations
\begin{equation}
    \begin{cases}
    \rho_1 (R, Z, \tau) +\Psi_1 (R, Z, \tau)=\rho_{m1} (Z, \tau) +\Psi_{m1} (Z, \tau),\\
    s_1 (R,Z,\tau)=s_{m1} (Z, \tau).
    \end{cases}
    \label{eq:Requil}
\end{equation}
Eq. (\ref{eq:Requil}) highlights that while the charge and potential change in the radial direction, the salt values remain constant.
The PNP equations \eqref{eq:PNPpert} with radial equilibrium and no Faradaic fluxes at the pore surface result in (see Refs. \cite{alizadeh2017multiscale,gupta2020charging,henrique2022charging}, and Ref. \cite{aslyamov2022analytical} for a related discussion of a formal perturbation analysis in the aspect ratio of symmetric electrolytes in slit pores)
\begin{equation}
    \begin{cases}
    \dfrac{\partial\rho_1}{\partial\tau}=(1+\beta\gamma)\left(\dfrac{\partial^2\rho_1}{\partial Z^2}+\dfrac{\partial^2\Psi_1}{\partial Z^2}\right)+\beta(1-\gamma^2)\dfrac{\partial^2s_1}{\partial Z^2},\\[10pt]
    \dfrac{\partial s_1}{\partial\tau}=\beta\left(\dfrac{\partial^2\rho_1}{\partial Z^2}+\dfrac{\partial^2\Psi_1}{\partial Z^2}\right)+(1-\beta\gamma)\dfrac{\partial^2s_1}{\partial Z^2}.
    \end{cases}
    \label{eq:pnpZ}
\end{equation}
Moreover, Poisson's equation in (\ref{eq:PNPpert}) can be simplified within the pore by asymptotically neglecting axial derivatives (for $\ell_p\gg a_p$), yielding
\begin{equation}
    -\dfrac{1}{R}\dfrac{\partial}{\partial R}\left(R\dfrac{\partial\Psi_1}{\partial R}\right)=\kappa^2\rho_1,
    \label{eq:poissonR}
\end{equation}
where we recall $\kappa=a_p/\lambda_D$ is the relative pore size, the nondimensional inverse Debye length. With the choice of applied potential as the small parameter in the perturbation expansion, we have $\Psi_1(R=1)=1$, in addition to the symmetry boundary condition $\partial\Psi_1/\partial R(R=0)=0$. Substituting the relation between charge density and potential given by Eq. (\ref{eq:Requil}) in Eq. (\ref{eq:poissonR}), and solving for the potential with the aforementioned boundary conditions, we find \cite{gupta2020charging,henrique2022charging}
\begin{equation}
    \dfrac{\Psi_1-\Psi_{m1}-\rho_{m1}}{1-\Psi_{m1}-\rho_{m1}}=\dfrac{I_0(\kappa R)}{I_0(\kappa)},
    \label{eq:poisson_sol}
\end{equation}
where $I_n$ is the modified Bessel function of the first kind. Evaluating this equation at $R=0$, we find
\begin{equation}
    \rho_{m1}=\dfrac{\Psi_{m1}-1}{I_0(\kappa)-1},
    \label{eq:rhopsi}
\end{equation}
and plugging this result back into Eq. (\ref{eq:poisson_sol}), we find the additional relation
\begin{equation}
    \Psi_1=\dfrac{I_0(\kappa)-I_0(\kappa R)}{I_0(\kappa)-1}\Psi_{m1}+\dfrac{I_0(\kappa R)-1}{I_0(\kappa)-1}.
\end{equation}
Using Eqs. (\ref{eq:poisson_sol}) and (\ref{eq:rhopsi}) in (\ref{eq:Requil}), we also find
\begin{equation}
    \rho_1=\rho_{m1}I_0(\kappa R).
\end{equation}

It should be noted that the same radial dependences of the variables as in a symmetric electrolyte are present \cite{henrique2022charging}, as expected, since the diffusivities and valences determine the coefficients of transport, which in turn set the axial transient behavior of the system, while the steady state remains the same. These known radial dependences promoted by radial equilibrium enable us to perform radial averages, defined for an arbitrary function $f(R)$ by
\begin{equation}
    \bar{f}=2\int_0^1f(R)R\,\mathrm{d}R.
\end{equation}
In fact, averaging the PNP Eqs. (\ref{eq:pnpZ}) over the cross section of the pore using Eqs. \eqref{eq:Requil} and \eqref{eq:poisson_sol} yields the effective transport equations
\begin{equation}
    \begin{cases}
    \dfrac{\partial\bar{\rho}_1}{\partial\tau}=\dfrac{1+\beta\gamma}{\tau_0}\dfrac{\partial^2\bar{\rho}_{1}}{\partial Z^2}+\beta(1-\gamma^2)\dfrac{\partial^2\bar{s}_{1}}{\partial Z^2},\\[10pt]
    \dfrac{\partial \bar{s}_{1}}{\partial\tau}=\dfrac{\beta}{\tau_0}\dfrac{\partial^2\bar{\rho}_{1}}{\partial Z^2}+(1-\beta\gamma)\dfrac{\partial^2\bar{s}_{1}}{\partial Z^2},
    \end{cases}
    \label{eq:pnpav}
\end{equation}
in the pore ($0<Z<1$), $\tau_0=2 I_1(\kappa)/\kappa I_0(\kappa)$ being the charging timescale of a symmetric electrolyte \cite{henrique2022charging}.

Eqs. \eqref{eq:pnpsdlav} and \eqref{eq:pnpav} summarize our effective model for ion transport in asymmetric electrolytes subjected to low applied potentials, a key result of our paper. In contrast to the symmetric electrolyte case, a diffusivity mismatch induces a concentration overpotential even at the low potential limit, coupling the charge dynamics to the salt evolution. Likewise, charge inhomogeneity produces salt variations, coupling the salt dynamics to the charge evolution. We illustrate the physical information in the model by contrasting two instructive limits. For overlapping double layers ($\kappa\ll 1$ and $\tau_0\sim 1$), Eqs. \eqref{eq:pnpsdlav} can be combined to yield the following dimensionless ion concentration equations: \begin{equation}
    \begin{cases}
    \dfrac{\partial\bar{c}_{+1}}{\partial\tau}=(1+\beta)\dfrac{\partial^2\bar{c}_{+1}}{\partial Z^2},\\[10pt] \dfrac{\partial\bar{c}_{-1}}{\partial\tau}=(1-\beta)\dfrac{\partial^2\bar{c}_{-1}}{\partial Z^2}.
    \end{cases}
    \label{eq:pnpthinEDL}
\end{equation}
In this regime, diffusion dominates over electromigration to set the effective axial ion transport timescales, dimensionally given by $\ell_p^2/D_\pm$ for cations and anions, respectively. In the thin double layer regime ($\kappa\gg 1$, $\tau_0\sim 2/\kappa$), we can rescale time by $\mathcal{T}=\tau/\tau_0$ to balance charge dynamics and charge-induced diffusion to obtain
\begin{equation}
    \begin{cases}
    \dfrac{\partial\bar{\rho}_1}{\partial\mathcal{T}}=(1+\beta\gamma)\dfrac{\partial^2\bar{\rho}_{1}}{\partial Z^2}+\dfrac{2\beta(1-\gamma^2)}{\kappa}\dfrac{\partial^2\bar{s}_{1}}{\partial Z^2},\\[10pt]
    \dfrac{\partial \bar{s}_{1}}{\partial\mathcal{T}}=\beta\dfrac{\partial^2\bar{\rho}_{1}}{\partial Z^2}+\dfrac{2(1-\beta\gamma)}{\kappa}\dfrac{\partial^2\bar{s}_{1}}{\partial Z^2},
    \end{cases}
    \label{eq:pnpoverEDL}
\end{equation}
Here, electromigration dominates diffusion and the presence of a concentration overpotential in addition to the Ohmic flux couples the charge and salt dynamics. However, for short times, since $1/\kappa\ll 1$, salt does not affect the charge dynamics, implying charging at a timescale $2\lambda_D\ell_p^2(z_+-z_-)/[(z_+D_+-z_-D_-)a_p]$ in dimensional terms. Remarkably, despite the thinness of the double layer, ambipolar diffusivity does not control axial salt or charge transport in this regime. Rather, the effective diffusivity for charge transport is governed by the electromigrative ionic mobilities through an arithmetic mean of the ionic diffusivities weighted by the respective valences, and the charge dynamics gives rise to a nontrivial salt dynamics.

As just discussed, this system has a rich dynamics with dependences on valence and diffusivity mismatches, and relative pore size, whose interpretations provide physical insights into the mechanics of pore charging.

% Lastly, we make the rescaling $T=\tau/\tau_0$ in Eqs. (\ref{eq:pnpsdl}) and (\ref{eq:pnpav}), which is useful in order to have a balanced charge conservation equation with $T=O(1)$ for all relative pore sizes:
% \begin{equation}
% \begin{cases}
%     (1+\beta\gamma)\dfrac{\partial^2\Psi_1}{\partial Z^2}+\beta(1-\gamma^2)\dfrac{\partial^2s_1}{\partial Z^2}=0,\\[10pt]
%     \dfrac{\partial s_1}{\partial T}=\dfrac{D_s(1-\beta^2)\tau_0}{D_p(1+\beta\gamma)}\dfrac{\partial^2s_1}{\partial Z^2},
% \end{cases}
% \label{eq:pnpsdlT}
% \end{equation}
% in the SDL and
% \begin{equation}
%     \begin{cases}
%     \dfrac{\partial\Psi_{m1}}{\partial T}=(1+\beta\gamma)\dfrac{\partial^2\Psi_{m1}}{\partial Z^2}+\dfrac{\beta(1-\gamma^2)}{\sigma}\dfrac{\partial^2s_{m1}}{\partial Z^2},\\[10pt]
%     \dfrac{\partial s_{m1}}{\partial T}=\beta\sigma\tau_0\dfrac{\partial^2\Psi_{m1}}{\partial Z^2}+(1-\beta\gamma)\tau_0\dfrac{\partial^2s_{m1}}{\partial Z^2}.
%     \end{cases}
%     \label{eq:pnpavT}
% \end{equation}
% in the pore. \textcolor{red}{These are important equations. Highlight the differences between this equation and the symmetric case. Describe what factors come in where, and how do these equations show what happens when the region is confined, i.e., the pore and when the region is unconfined, i.e., the SDL. Along the same lines, perhaps discussion of $\kappa \gg 1$ would should salt reducing to a simple case of ambipolar diffusivity should also be mentioned here...}

\subsubsection{Boundary and Initial Conditions}
\label{sec:IC}

Now we state the boundary and initial conditions of the reduced-order model. The reservoir is characterized by properties unaffected by the applied field, such that
\begin{equation}
\bar{s}_{1}(Z=-\ell_s/\ell_p,\tau)=0.
\label{eq:bulksalt}
\end{equation}
We ascribe a reference average potential to this region,
\begin{equation}
\bar{\Psi}_{1}(Z=-\ell_s/\ell_p,\tau)=0.
\label{eq:bulkpot}
\end{equation}
We neglect the occurrence of leakage or reactions on any surfaces of the pore, thus
\begin{equation}
\left.\dfrac{\partial\bar{\Psi}_{1}}{\partial Z}\right|_{Z=1}=\left.\dfrac{\partial \bar{s}_{1}}{\partial Z}\right|_{Z=1}=0.
\label{eq:blockelec}
\end{equation}
Following previous works \cite{gupta2020charging,henrique2022charging}, we include in the effective one-dimensional model a transition region connecting the electroneutral SDL to the inside of a charged pore of arbitrary relative size, where significant changes in charge density and electric potential may occur. The mismatch in the lengths of the double layer and the pore, $\lambda_D/\ell_p\ll  1$, ensures that this transition region is thin compared to the pore length, and we follow Refs. \cite{gupta2020charging,henrique2022charging} in neglecting the charge and salt fluxes in the transition region in order to satisfy charge balance, and approximating them by Newton quotients. This approximation implies
\begin{equation}
    \bar{\rho}_{1,\mathrm{right}}+\bar{\Psi}_{1,\mathrm{right}}=\bar{\Psi}_{1,\mathrm{left}},
\end{equation}
where the index ``left'' refers to the SDL side of the transition $(Z=0^-)$, and ``right'' to the pore side of it $(Z=0^+)$; see Fig. \ref{fig:pore}. Averaging the radial equilibrium relation between charge density and potential in the pore in the form of Eq. \eqref{eq:poisson_sol}, we get
\begin{equation}
    \bar{\Psi}_{1,\mathrm{left}}=1+\dfrac{\bar{\rho}_{1,\textrm{right}}}{\tau_0}.
    \label{eq:psi_rl}
\end{equation}
Likewise, the neglect of charge and salt fluxes in the transition region also implies the continuity of salt density across it,
\begin{equation}
    \bar{s}_{1,\mathrm{left}}=\bar{s}_{1,\mathrm{right}}.
    \label{eq:s_rl}
\end{equation}
Moreover, we impose charge conservation between the ends of the transition region with negligible charge storage by matching the charge and salt currents on both ends, $\alpha \bar{J}_Z|_{Z=0^-}=\bar{J}_Z|_{Z=0^+}$ and $\alpha \bar{W}_Z|_{Z=0^-}=\bar{W}_Z|_{Z=0^+}$, and linear combinations of these equations yield
\begin{equation}
\begin{cases}
    \alpha\left.\dfrac{\partial\bar{\Psi}_{1}}{\partial Z}\right|_{Z=0^-}=\dfrac{1}{\tau_0}\left.\dfrac{\partial\bar{\rho}_{1}}{\partial Z}\right|_{Z=0^+},\\[10pt]
    \alpha\left.\dfrac{\partial \bar{s}_{1}}{\partial Z}\right|_{Z=0^-}=\left.\dfrac{\partial \bar{s}_{1}}{\partial Z}\right|_{Z=0^+},
\end{cases}
\label{eq:bcs_cnsv}
\end{equation}
where $\alpha=A_sD_s/(A_pD_p)$ (see Eqs. \eqref{eq:PNPpert} and the radial equilibrium Eqs. \eqref{eq:poisson_sol} and \eqref{eq:rhopsi} for the expressions of the radially averaged fluxes). We note that while this charge transport condition is governed by an analogue of a Biot number $\mathrm{Bi}=A_sD_s\ell_p/(A_pD_p\ell_s)$ for symmetric electrolytes at low potentials \cite{biesheuvel2010nonlinear,gupta2020charging,henrique2022charging}, and the potential profile in the SDL of an asymmetric electrolyte is not necessarily linear, such that the ratio of lengths does not appear explicitly in these boundary conditions.

Initially, the salt and charge densities have not yet been affected by the applied field, therefore their initial corrections due to the applied potential are absent, i.e., $\bar{s}_{1}(Z,\tau=0)=0$ over both the pore and SDL regions. On the other hand, a potential field instantaneously develops due to the presence of the pore surface potential. With electroneutrality initially, Laplace's equation holds everywhere for the potential. Thus, the initial potential field follows to be
\begin{equation}
    \bar{\Psi}_{1}(Z,T=0)=\begin{cases}
    \dfrac{\ell_p}{\ell_s}Z+1,\quad -\dfrac{\ell_s}{\ell_p}\le Z<0,\\[10pt]
    1,\quad\quad\quad 0\le Z\le 1.
    \end{cases}
    \label{eq:psi0}
\end{equation}
Eqs. (\ref{eq:pnpsdlav}) and (\ref{eq:pnpav}) with the boundary and initial conditions (\ref{eq:bulksalt})--(\ref{eq:psi0}) have the symmetry $\bar{\rho}_{1}(\tau;\beta,\gamma)=\bar{\rho}_{1}(\tau;-\beta,-\gamma)$, $\bar{\Psi}_{1}(\tau;\beta,\gamma)=\bar{\Psi}_{1}(\tau;-\beta,-\gamma)$, $\bar{s}_{1}(\tau;\beta,\gamma)=-\bar{s}_{1}(\tau;-\beta,-\gamma)$. We remark that in the regime of high SDL conductance, $\alpha\to \infty$, Eq. (\ref{eq:bcs_cnsv}) simplifies in such a way that the dynamics of the SDL become trivial, i.e., $\bar{\Psi}_1(Z,\tau)\to 0$ and $\bar{s}_1(Z,\tau)\to 0$ for $-\ell_s/\ell_p<Z<0$. We discuss the physics of this scenario in greater detail in Sec. \ref{sec:soln}.
\begin{figure}[t!]
    \centering
    \includegraphics[max width=\textwidth]{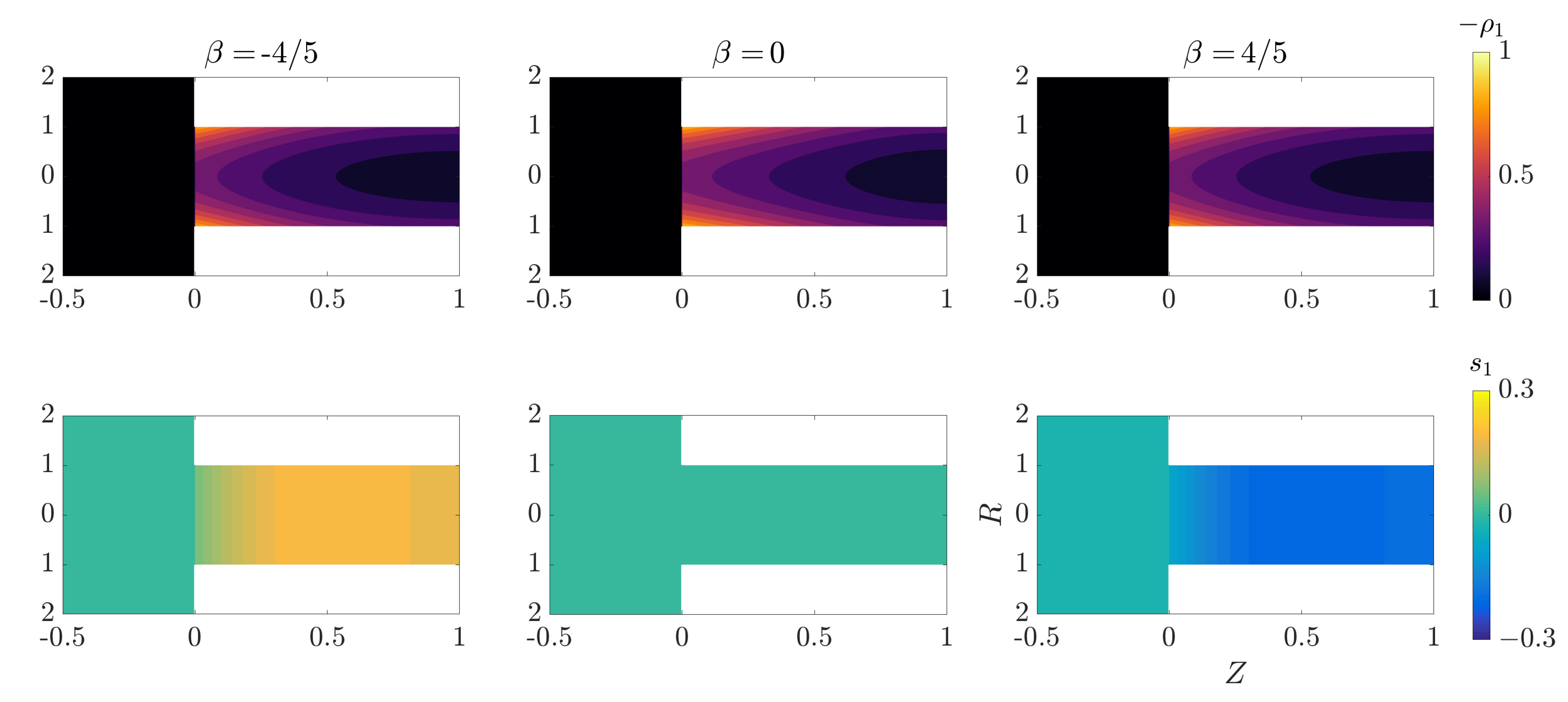}
    \caption{\textbf{Predictions of the perturbation model}. Contour plots of the charge (upper row) and salt (lower row) densities found from the reduced-order model described by Eqs. \eqref{eq:pnpsdlav} and \eqref{eq:pnpav} for $\tau=0.2$, $\alpha=4$, $\gamma=0$, and $\kappa=2$. All plots share the same axes. A mismatch in diffusivities induces a nontrivial salt dynamics coupled to the charge dynamics, mildly altering the charging timescale of the pore.}
    \label{fig:contour}
\end{figure}

\section{Numerical Results}\label{sec:num}

In this section, we discuss the results of our linearized model, assess its validity and discuss some physical implications of its solution. We solve the averaged linearized Eqs. (\ref{eq:pnpsdlav}) and (\ref{eq:pnpav}) by a compact fourth-order finite-difference scheme \cite{gamet1999compact}, and the full Eqs. (\ref{Eq: pnp}) and (\ref{Eq: fluxes}) by direct numerical simulations (DNS) \cite{weller1998tensorial,jasak2007openfoam} (see Ref. \cite{gupta2020charging} for the details of the DNS). The charge and salt profiles found from the finite-difference integration of our reduced-order equations is illustrated in Fig. \ref{fig:contour}. The key ingredient of this asymmetric analysis is the inclusion of salt migration due to mismatch in diffusivities and valences, which can lead to significant salt intake or depletion within the pore. Salt transport both within and outside of the pore is also coupled to the charging process, mildly influencing the charging dynamics. As will be discussed in Sec. \ref{sec:soln} and \ref{Sec:app}, simultaneous asymmetries in valences and diffusivities may produce a more significant impact on the charging dynamics.

For simplicity, we compare the results of both methods for the centerline values of charge and salt densities, and electric potential. We note that in the pore, these properties are related to the cross-sectional averages we find from Eqs. \eqref{eq:Requil} and \eqref{eq:poisson_sol} by
\begin{equation}
    \rho_{m1}=\dfrac{\bar{\rho}_1}{\tau_0I_0(\kappa)},
    \label{eq:rho_bar_m}
\end{equation}
\begin{equation}
    \Psi_{m1}=\dfrac{I_0(\kappa)-1}{\tau_0I_0(\kappa)}\bar{\rho}_1+1,
    \label{eq:psi_bar_m}
\end{equation}
and
\begin{equation}
    s_{m1}=\bar{s}_1.
    \label{eq:s_bar_m}
\end{equation}

\subsection{Effects of Valence Contrast}
\begin{figure*}[t!]
    \centering
    \begin{subfigure}{0.33\textwidth}
    \centering
    \caption{}
    %\phantom{\large{$\gamma=0$}}\par\medskip
    \includegraphics[max width=.95\textwidth]{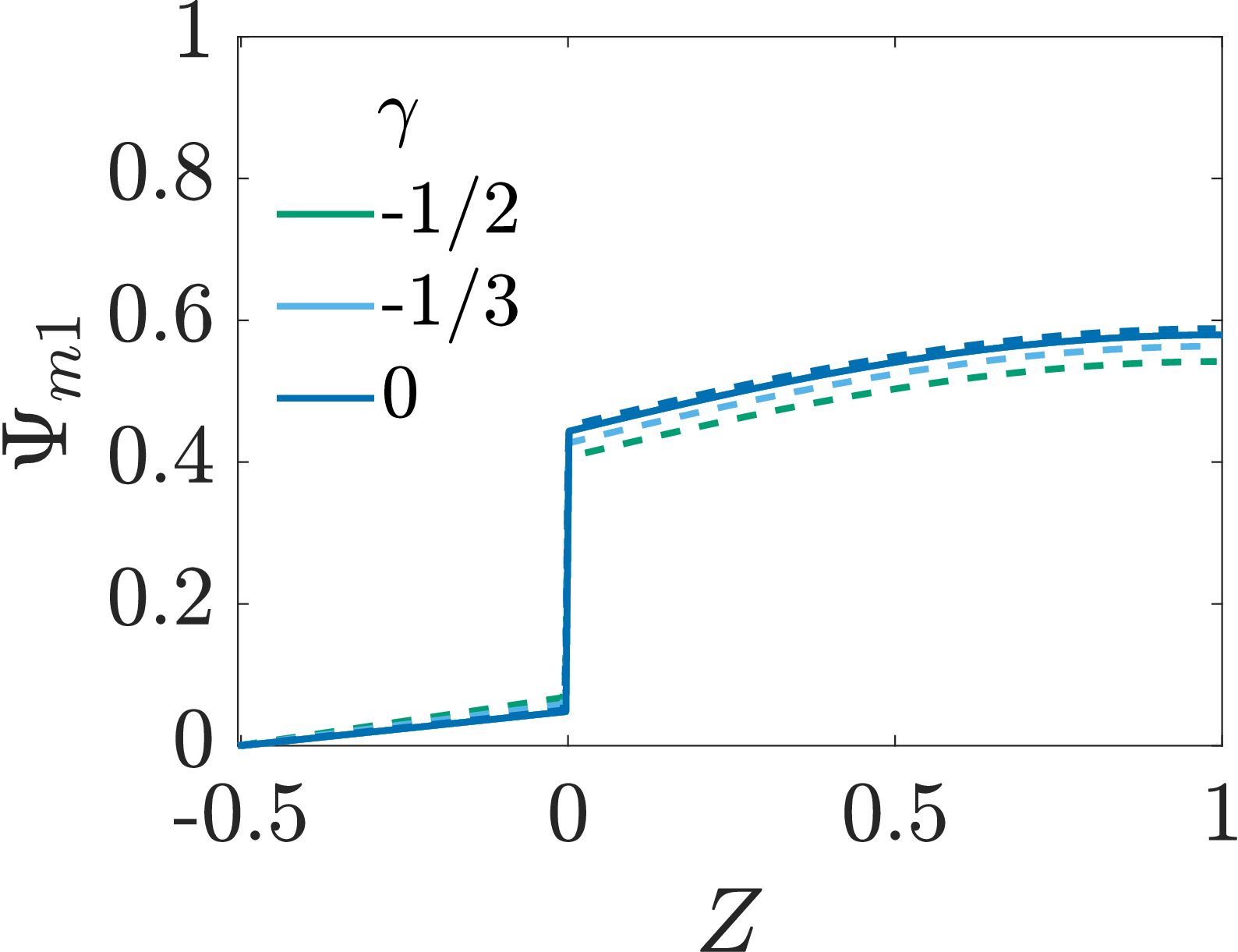}
    \end{subfigure}%
    \begin{subfigure}{0.33\textwidth}
    \centering
    \caption{}
    %\large{$\gamma=0$}\par\medskip
    \includegraphics[max width=.95\textwidth]{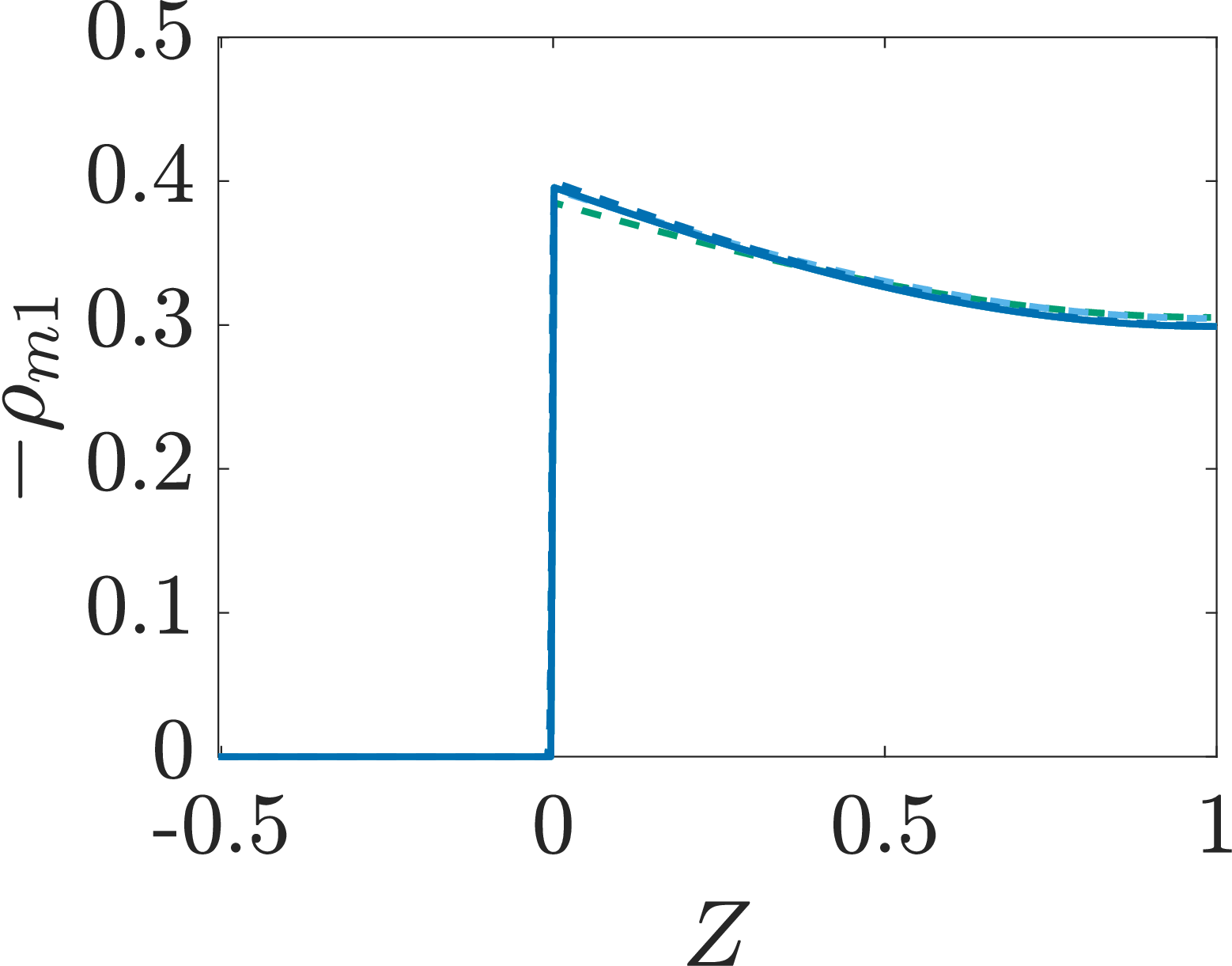}
    \end{subfigure}%
    \begin{subfigure}{0.33\textwidth}
    \caption{}
    %\phantom{\large{$\gamma=0$}}\par\medskip
    \centering
    \includegraphics[max width=.95\textwidth]{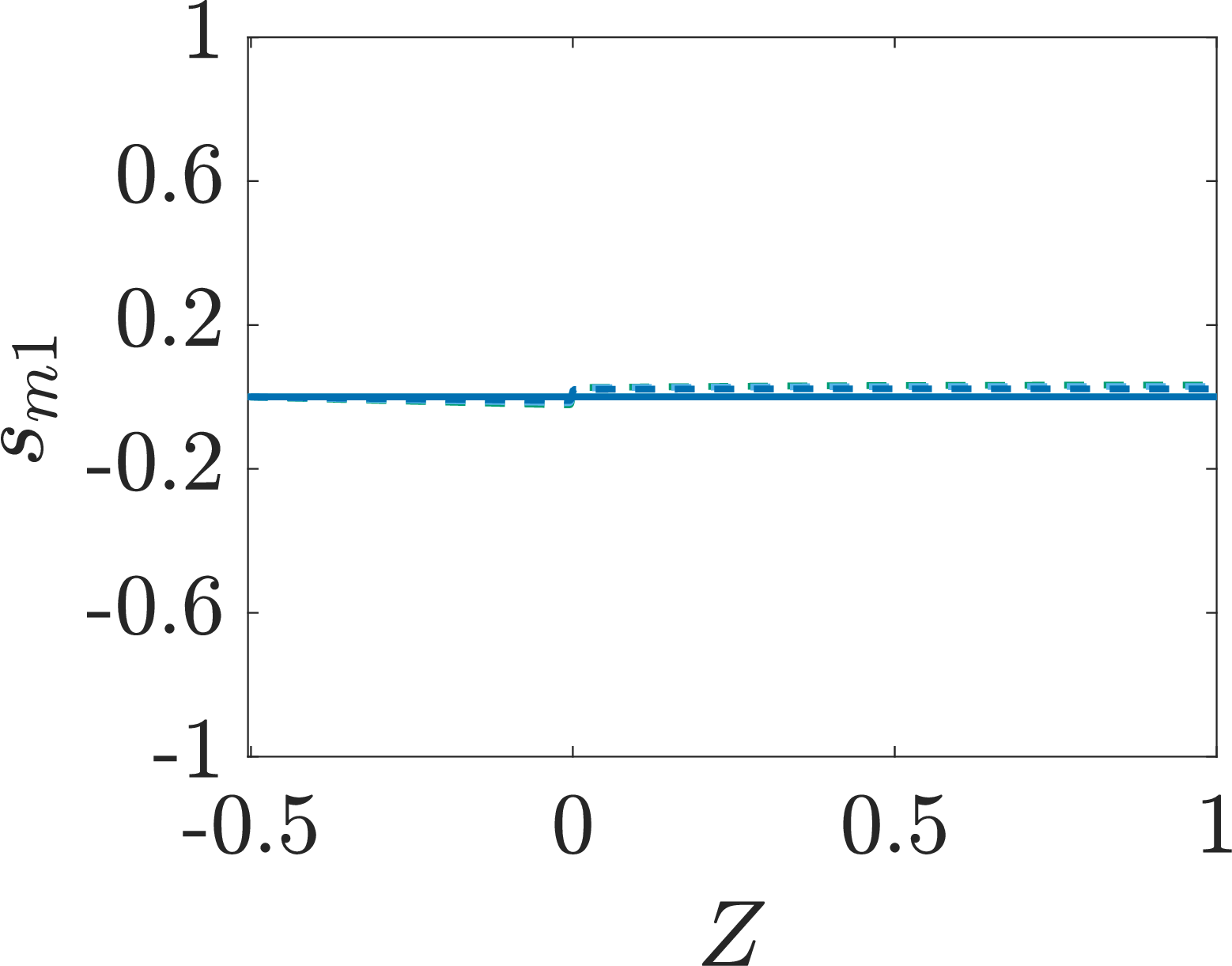}
    \end{subfigure}
    \caption{\textbf{Comparison between the perturbation model and DNS for different ion valences}. Centerline profiles of the independent variables at $tD_\pm/\ell_p^2=0.52$ for a cation valence $z_+=1$ and different anion valences: $z_-=-1$ $(\gamma=0)$, $z_-=-2$ $(\gamma=-1/3)$ and $z_-=-3$ $(\gamma=-1/2)$. Solutions of the reduced-order model Eqs. \eqref{eq:pnpsdlav} and \eqref{eq:pnpav} represented by solid lines, and DNS of Eqs. \eqref{Eq: pnp} and \eqref{Eq: fluxes} by dashed lines. a) electric potential, b) charge density, c) salt density. For each valence, the reference concentration $c_0$ is adjusted such that $\kappa=2$. $\beta=0$, $\alpha=4$, and $e\phi_D/(k_BT)=0.4$ for all plots. A contrast in valences only produces a variation in the Debye length, with no impact of the response of the linearized system when the relative pore size $\kappa$ is kept constant. The reduced-order model shows good agreement with the DNS.}
    \label{fig:profilesgammaZ}
\end{figure*}

The effects of valence contrast are shown in Fig. \ref{fig:profilesgammaZ} for a constant relative pore size $\kappa=2$. In this limit, with no contrast in diffusivities ($\beta=0$), Eqs. (\ref{eq:pnpsdlav}) and (\ref{eq:pnpav}) do not change. Essentially, the effect of a contrast in valences is to change the ionic strength of the electrolyte, and consequently its Debye length. Adjusting the reference concentration in such a way that the relative pore size remains constant, a contrast in valences alone does not change the potential, charge, or salt profiles in the linear regime $|\Psi_D|\ll 1$, i.e., salt density is approximately constant in this scenario and the results in Refs. \cite{gupta2020charging,henrique2022charging} hold as long as the correction in relative pore size induced by the effect of ionic strength over the Debye length is taken into account. There is good agreement between the results of the linearized model and the DNS, with the greatest quantitative discrepancy of about 7\% occurring for potential when $\gamma=-1/2$, and even in this case charge density shows good agreement. The comparison drawn in Fig. \ref{fig:profilesgammaZ} is effected with a concomitant variation in the average valence $z_\textrm{avg}$ at a constant applied potential $e\phi_D/(k_BT)=0.4$, such that higher contrasts fall further from the strict domain of validity of the asymptotic approximation $|\Psi_D|\ll 1$. Non-linear effects might become more important, but the qualitative and quantitative results in the plots agree well in the cases analysed.

\subsection{Effects of Diffusivity Contrast}

\begin{figure*}[t!]
    \centering
    \begin{subfigure}{0.33\textwidth}
    \centering
    \caption{}
    %\phantom{\large{$\gamma=0$}}\par\medskip
    \includegraphics[max width=.95\textwidth]{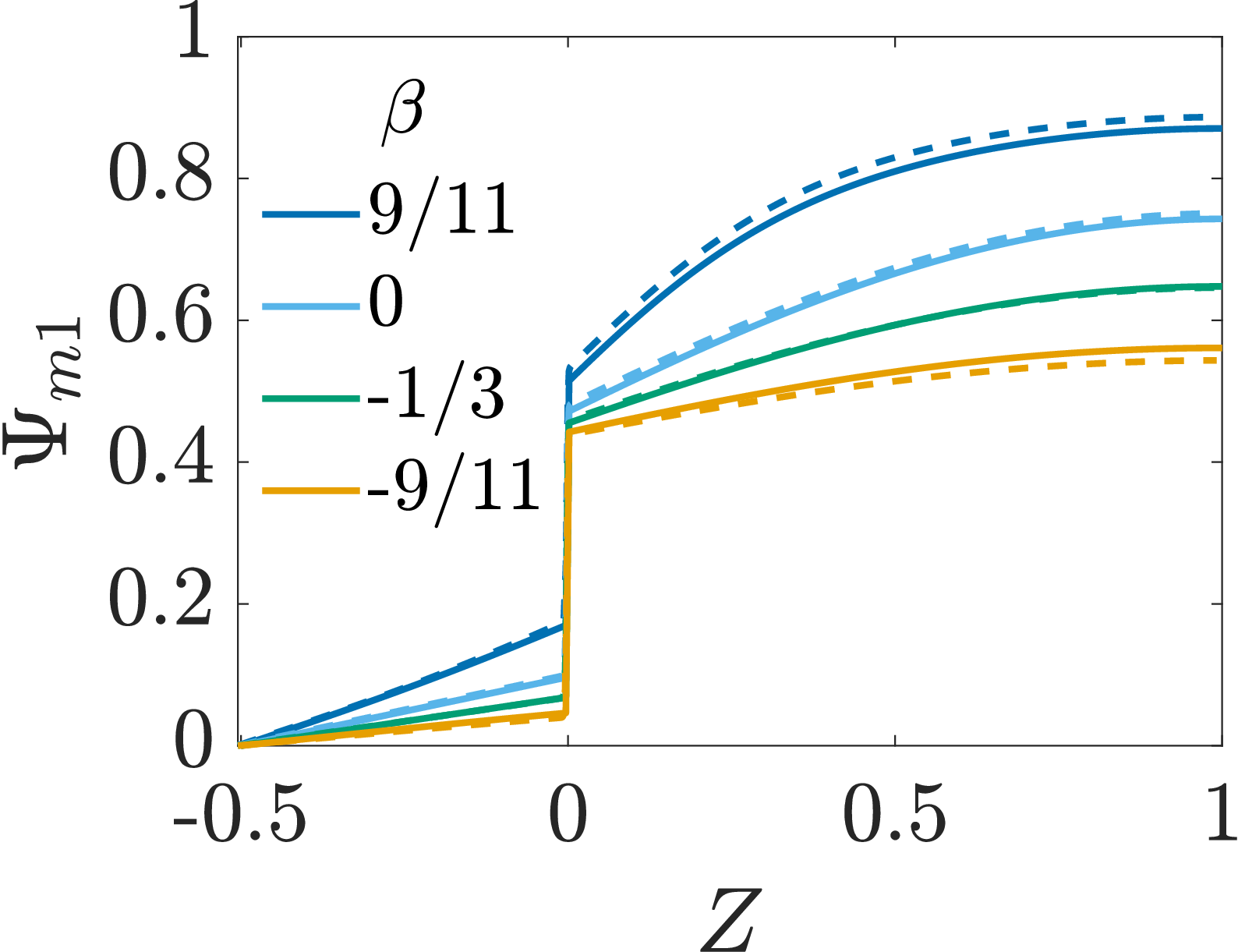}
    \end{subfigure}%
    \begin{subfigure}{0.33\textwidth}
    \centering
    \caption{}
    %\large{$\gamma=0$}\par\medskip
    \includegraphics[max width=.95\textwidth]{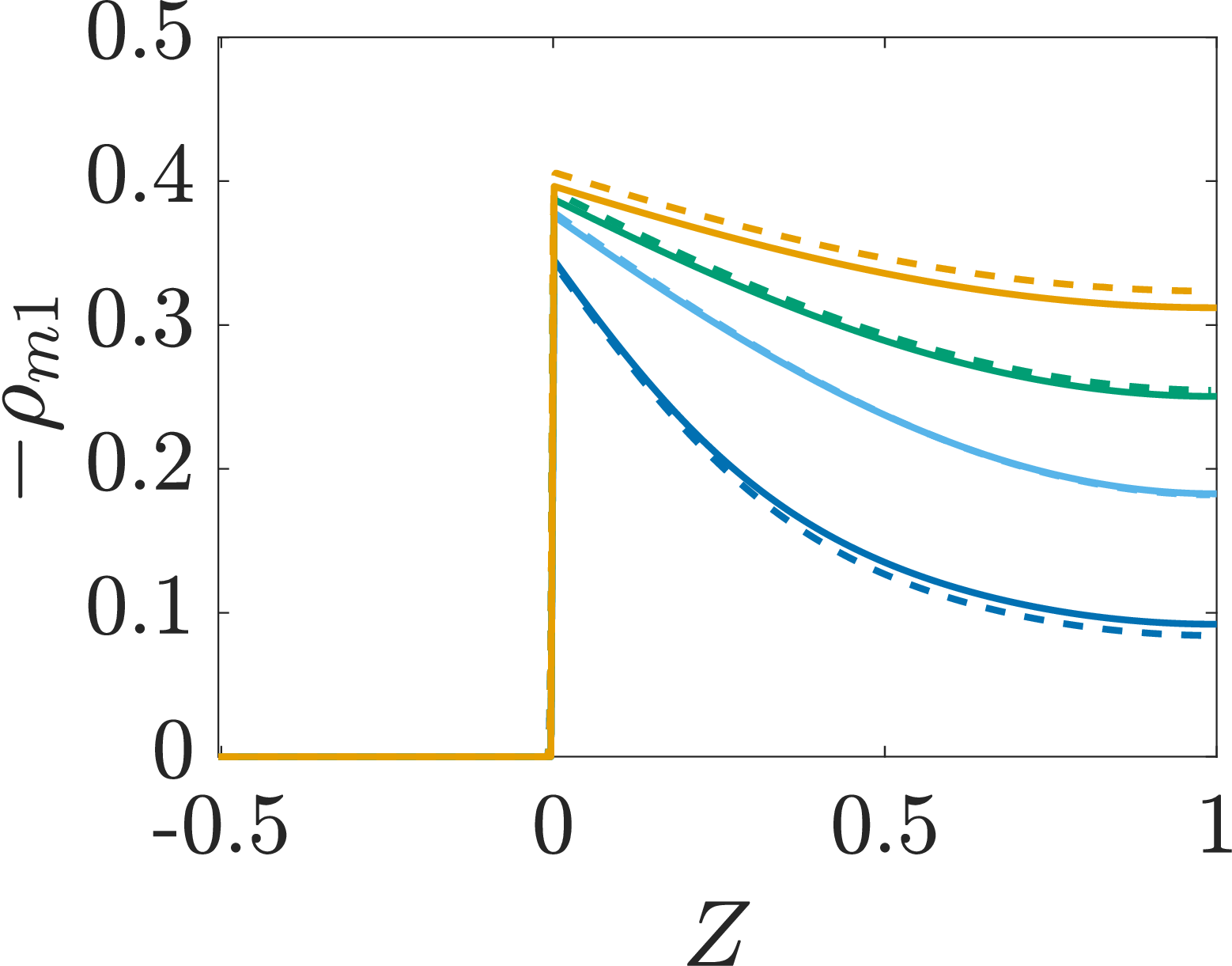}
    \end{subfigure}
    \begin{subfigure}{0.33\textwidth}
    \caption{}
    %\phantom{\large{$\gamma=0$}}\par\medskip
    \centering
    \includegraphics[max width=.95\textwidth]{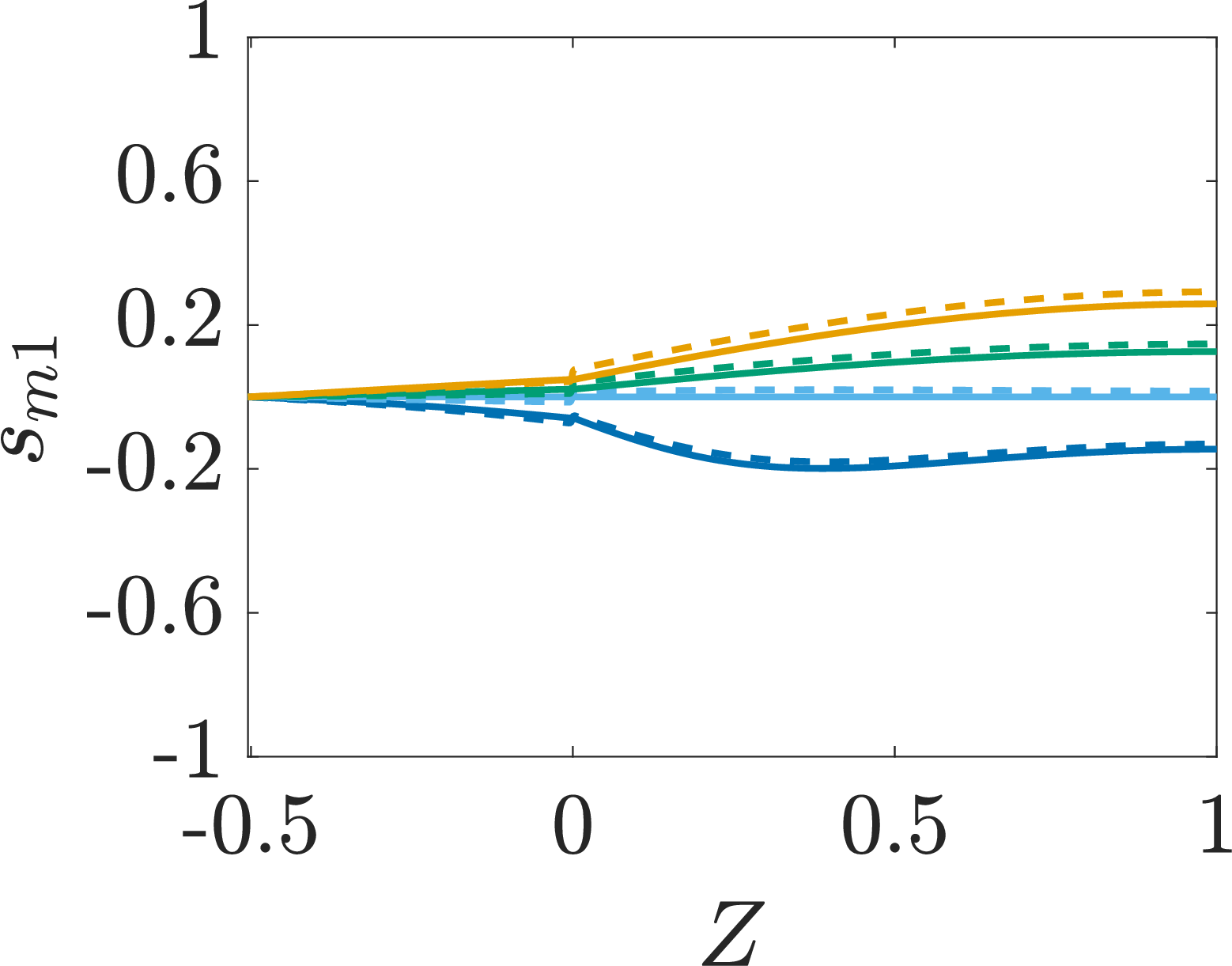}
    \end{subfigure}
    \caption{\textbf{Comparison between the perturbation model and DNS for different ion diffusivities}. Centerline profiles of the independent variables at $tD_+/\ell_p^2=0.28$ for different values of anion diffusivity: $D_-/D_+=0.1$ ($\beta=9/11$), $D_-/D_+=1$ ($\beta=0$), $D_-/D_+=2$ ($\beta=-1/3$) and $D_-/D_+=10$ ($\beta=-9/11$). Solutions of the reduced-order model Eqs. \eqref{eq:pnpsdlav} and \eqref{eq:pnpav} represented by solid lines, and DNS of Eqs. \eqref{Eq: pnp} and \eqref{Eq: fluxes} by dashed lines. a) potential, b) charge density, c) salt density. $\gamma=0$, $\alpha=4$, and $e\phi_D/(k_BT)=0.4$ for all plots. The reduced-order model shows good agreement with the DNS. A contrast in diffusivities accelerates or hinders charging, in addition to producing appreciable salt enrichment or depletion in the SDL and pore regions.}
    \label{fig:profilesbetaZ}
\end{figure*}

Next, we examine the results in the presence of a contrast in diffusivities, but not in valences.  Fig. \ref{fig:profilesbetaZ} shows that the potential screening and the charge buildup at the centerline at a given time are notably dependent on the ionic diffusivities, i.e., they alter the charging timescale. Furthermore, it produces salt accumulation or depletion in the pore, with a non-monotonic profile at short times. This salt dynamics is a key feature of asymmetric diffusivities, not observed in previous works on symmetric electrolytes at low potentials \cite{gupta2020charging,henrique2022charging}. We note that even in the case of asymmetric diffusivities, the linearized model presents good agreement with the DNS for anion diffusivities that vary by two orders of magnitude as $D_-$ goes from $0.1D_+$ to $10D_+$. The cases shown in Fig. \ref{fig:profilesbetaZ} have an intermediate relative pore size, $\kappa=2$, where neither diffusion nor migration completely dominates ion transport. In this scenario, the charging timescale is neither given by a valence-weighted average of the diffusivities, nor by the slowest of the diffusion times of the ions. 
The ionic diffusivities affect the equations describing the transient process in two ways; first, they set the reference timescale $\ell_p^2/D_p$ based on the arithmetic mean of the diffusivities to normalize the solutions. Physically, this corresponds to a change in the overall rate of the charging process by controlling the rate of diffusion in an average sense. When the arithmetic mean of the diffusivities is increased, with all else constant, this effective diffusion coefficient is boosted, resulting in commensurately faster transport for both charge and salt. Second, the absence of a double layer in the initial condition imposes a linear electric field in the static diffusion layer, with equal concentrations of cations and anions. Under this imposed field in the SDL, the asymmetric electromigrative mobilities of cations and anions produce a contrast in their rates of transport, producing salt depletion or intake, promoting salt-induced charge and salt fluxes, and modifying the charge, salt, and potential profiles. The dynamics of the overall process can also be dependent on the coupling to the SDL, with a characteristic timescale based on the ambipolar diffusivity. The physical interpretation of charging dependence on electrolyte asymmetry is further explored in Sec. \ref{sec:soln}, where we simplify to model in the limit of high SDL conducitivity to obtain additional physical insights.

\section{An Analytical Solution for High SDL Conductance}\label{sec:soln}

Though the SDL is a feature of the model that enables us to include the resistance to ion transport outside the pore, the system of Eqs. (\ref{eq:pnpav}) coupled to the SDL through the contact boundary conditions (\ref{eq:psi_rl}) to (\ref{eq:bcs_cnsv}) is challenging to solve analytically. However, in the limit $\alpha\to\infty$, we can obtain significant insights of all the parameter dependences via an analytical solution of the linearized equations. In this limit, corresponding to high SDL conductance, Eqs. (\ref{eq:bcs_cnsv}) simplify to
\begin{equation}
    \begin{cases}
    \left.\dfrac{\partial\bar{\Psi}_{1}}{\partial Z}\right|_{Z=0^-}=0,\\[10pt]
    \left.\dfrac{\partial \bar{s}_{1}}{\partial Z}\right|_{Z=0^-}=0,
    \end{cases}
\end{equation}
allowing for the trivial solution of the SDL Eqs. \eqref{eq:pnpsdlav}. Therefore, in this setting, the transition region jump conditions (\ref{eq:psi_rl}) and (\ref{eq:s_rl}) are satisfied with $\bar{s}_{1,\mathrm{left}}=0$ and $\bar{\Psi}_{1,\mathrm{left}}=0$. 

In this case, we rescale time by $\mathcal{T}=\tau/\tau_0$ for a maximal balance between charge dynamics and charge-induced diffusion, and define the vector
\begin{equation}
    \mathbf{f}=\begin{bmatrix}
    \bar{\rho}_{1}+\tau_0\\[5pt]
    \bar{s}_{1}
    \end{bmatrix}
    \label{eq:defphi}
\end{equation}
and the matrix
\begin{equation}
    A=\begin{bmatrix}
    1+\beta\gamma & \beta(1-\gamma^2)\tau_0\\[5pt]
    \beta & (1-\beta\gamma)\tau_0
    \end{bmatrix}
\end{equation}
to recast the problem described by Eqs. \eqref{eq:pnpav} and \eqref{eq:bulksalt}--\eqref{eq:psi0} in the form of the vector partial differential equation
\begin{equation}
    \dfrac{\partial\mathbf{f}}{\partial \mathcal{T}}=A\dfrac{\partial^2\mathbf{f}}{\partial Z^2}
    \label{eq:highalpha}
\end{equation}
with boundary conditions
\begin{equation}
    \begin{cases}
    \mathbf{f}(Z=0^+,\mathcal{T})=0,\\[5pt]
    \left.\dfrac{\partial\mathbf{f}}{\partial Z}\right|_{Z=1}=0,
    \end{cases}
    \label{eq:BCs_f}
\end{equation}
and initial condition
\begin{equation}
    \mathbf{f}(Z,\mathcal{T}=0)=\begin{bmatrix}
    \tau_0\\[5pt]
    0
    \end{bmatrix}.
    \label{eq:IC_f}
\end{equation}
It should be noted that the inclusion of $\tau_0$ in the definition of $\mathbf{f}$ (Eq. \eqref{eq:defphi}) ensures the homogeneity of the boundary conditions. The initial value problem given by Eqs. \eqref{eq:highalpha} and \eqref{eq:IC_f} has the formal solution
\begin{equation}
    \mathbf{f}(Z,\mathcal{T})=\exp\left(\mathcal{T} A\dfrac{\partial^2}{\partial Z^2}\right)\mathbf{f}(Z,\mathcal{T}=0).
    \label{eq:formal_sol}
\end{equation}
Next, in order to satisfy the boundary conditions, we represent the initial condition in a Fourier series,
\begin{equation}
    \mathbf{f}(Z,\mathcal{T}=0)=2\tau_0\sum_{n=1}^\infty \dfrac{\sin(\xi_nZ)}{\xi_n}\begin{bmatrix} 1\\
    0
    \end{bmatrix},
    \label{eq:fourier_ic}
\end{equation}
where $\xi_n=(2n-1)\pi/2$, $n\in\mathbb{N}$ are the eigenvalues of the Sturm-Liouville operator $\mathcal{L}=\partial^2/\partial Z^2$ with BCs \eqref{eq:BCs_f}. Note that upon the application of the exponential operator in Eq. (\ref{eq:formal_sol}), as per its Taylor series definition, each second derivative series produces a multiplicative term $-\xi_n^2$, such the substitution of Eq. (\ref{eq:fourier_ic}) in (\ref{eq:formal_sol}) gives
\begin{equation}
     \mathbf{f}=2\tau_0\sum_{n=1}^\infty \dfrac{\sin(\xi_nZ)}{\xi_n}\exp(-\xi_n^2 \mathcal{T}A)\begin{bmatrix} 1\\
    0
    \end{bmatrix}.
    \label{Eq:phihighalpha}
\end{equation}
The coupling between potential and salt density appears through the exponential of the coefficient matrix. By Sylvester's formula, the exponential of a $k\times k$ matrix $\mathcal{A}$ can be rewritten in terms of its eigenvalues $\lambda_i$ as \cite{horn1994topics}
\begin{equation}
    \exp(\mathcal{A})=\sum_{i=1}^k\exp(\lambda_i)\prod_{\substack{j=1\\j\ne i}}^k\dfrac{1}{\lambda_i-\lambda_j}(\mathcal{A}-\lambda_j\mathbb{1}),
    \label{eq:sylvester}
\end{equation}
where $\lambda_i$ are the eigenvalues of $\mathcal{A}$ and $\mathbb{1}$ is the identity matrix. In our problem, we use Eq. (\ref{eq:sylvester}) for the $2\times 2$ matrix $\mathcal{A}=-\xi_n^2\mathcal{T}A$, yielding
\begin{equation}
\exp(-\xi_n^2\mathcal{T}A)=\exp(-\xi_n^2\mathfrak{I}_1\mathcal{T})\left[\left(\cosh(\xi_n^2\mathfrak{I}_2\mathcal{T})+\mathfrak{I}_1\dfrac{\sinh(\xi_n^2\mathfrak{I}_2\mathcal{T})}{\mathfrak{I}_2}\right)\mathbb{1}-\dfrac{\sinh(\xi_n^2 \mathfrak{I}_2\mathcal{T})}{\mathfrak{I}_2}A\right],
\label{Eq:expmat}
\end{equation}
where the $\mathfrak{I}_1$ and $\mathfrak{I}_2$ are invariants of $A$ defined by
\begin{equation}
    \mathfrak{I}_1=\dfrac{\tr A}{2}
\end{equation}
and
\begin{equation}
    \mathfrak{I}_2=\sqrt{-\det(A-\mathfrak{I}_1\mathbb{1})},
\end{equation}
From the form of the exponential given in Eq. (\ref{eq:sylvester}), since the eigenvalues of $A$ are given by
\begin{equation}
    \lambda=\dfrac{1}{2}\left\{1+\beta\gamma+(1-\beta\gamma)\tau_0\pm\sqrt{[1+\beta\gamma+(1-\beta\gamma)\tau_0]^2-4(1-\beta^2)\tau_0^2}\right\}
    \label{eq:eigA},
\end{equation}
we note that the solution (\ref{Eq:phihighalpha}) presents biexponential decay, with the two distinct timescales corresponding to the inverses of the eigenvalues. Aslyamov and Janssen \cite{aslyamov2022analytical} found similar coupled transient diffusion equations for nonlinear slit pore charging for thin double layers and symmetric diffusivities at high potentials, close to steady state, and used Sylvester's formula in that context to find biexponential decay on the order of $\ell_p^2/D_\pm=\ell_p^2/D_p$, which slows down long-time charging for thin double layers.

Eq. \eqref{eq:eigA} determines  the interplay of effects of pore size, controlling the diffusion-electromigration balance through $\tau_0$, and mismatch in diffusive and electromigrative mobilities of the ions, determined by the ionic valences and diffusivities, to set the distinct charging timescales in the pore. In the overlapping double layer regime $\kappa\ll 1$, yielding $\tau_0\sim 1$,
\begin{equation}
    \lambda\sim 1\pm\beta,
\end{equation}
whose inverses correspond dimensionally to $t\sim \ell_p^2/D_\pm$. The timescales in this limit are physically interpretable as the diffusion times of cations and anions. In this case, overlapping double layers present long-time charging hindered by the slowest process, be it diffusive expulsion of coions or attraction of counterions. At moderate relative pore sizes, both diffusive and electromigrative fluxes play a role in transport, such that a more complex interplay of valences and diffusivities and relative pore size sets the slow timescale. 

The final expressions for charge and salt densities and electric potential are found by substituting Eq. (\ref{Eq:phihighalpha}) and (\ref{Eq:expmat}) into Eq. (\ref{eq:defphi}).
For convenience, let us define the shorthands
\begin{equation}
    g_n(\mathcal{T})=\left[\cosh(\xi_n^2\mathfrak{I}_2\mathcal{T})+\dfrac{\mathfrak{I}_1-A_{11}}{\mathfrak{I}_2}\sinh(\xi_n^2\mathfrak{I}_2\mathcal{T})\right]\exp(-\xi_n^2\mathfrak{I}_1\mathcal{T})
    \label{eq:def_gn}
\end{equation}
and
\begin{equation}
    h_n(\mathcal{T})=-\dfrac{A_{21}}{\mathfrak{I}_2}\sinh(\xi_n^2\mathfrak{I}_2\mathcal{T})\exp(-\xi_n^2\mathfrak{I}_1\mathcal{T}),
    \label{eq:def_hn}
\end{equation}
where $A_{ij}$ are the matrix coefficients of $A$, to calculate charge and salt densities in the limit of high SDL conductance as
\begin{equation}
    \bar{\rho}_{1}(Z,\mathcal{T})=\tau_0\left(-1+2\sum_{n=1}^\infty \dfrac{\sin(\xi_nZ)g_n(\mathcal{T})}{\xi_n}\right)
    \label{eq:rhobar1nosdl}
\end{equation}
and
\begin{equation}
    \bar{s}_{1}(Z,\mathcal{T})=2\tau_0\sum_{n=1}^\infty \dfrac{\sin(\xi_nZ)h_n(\mathcal{T})}{\xi_n}.
    \label{eq:sbar1nosdl}
\end{equation}
Once the cross-sectional average quantities are known, radial equilibrium in the form of Eqs. \eqref{eq:Requil}, \eqref{eq:poisson_sol} and \eqref{eq:rho_bar_m}--\eqref{eq:s_bar_m} provides the spatial dependence in the pore. A few remarks on physical interpretations of the solution are in order. At steady state, $\bar{\rho}_1\to -\tau_0$ and $\bar{s}_1\to 0$ as $\mathcal{T}\to \infty$, indicating that the total charge storage capacity of the pore is set by the relative pore size through $\tau_0$ and is independent of electrolyte asymmetry. Furthermore, no permanent salt depletion or intake occurs in the low potential limit. The salt profile returns to its initial value at steady state. As expected, the sign of transient spatial changes is the salt profile is governed by the diffusivity mismatch through the coefficient $A_{21}=\beta$. If counterions are attracted faster than cations are expelled, salt is enriched in the pore. In the opposite case, salt is depleted in the pore.

\begin{figure*}[t]
    \centering
    \begin{subfigure}{0.33\textwidth}
    \centering
    \caption{}
    %\phantom{\large{$\gamma=0$}}\par\medskip
    \includegraphics[max width=.95\textwidth]{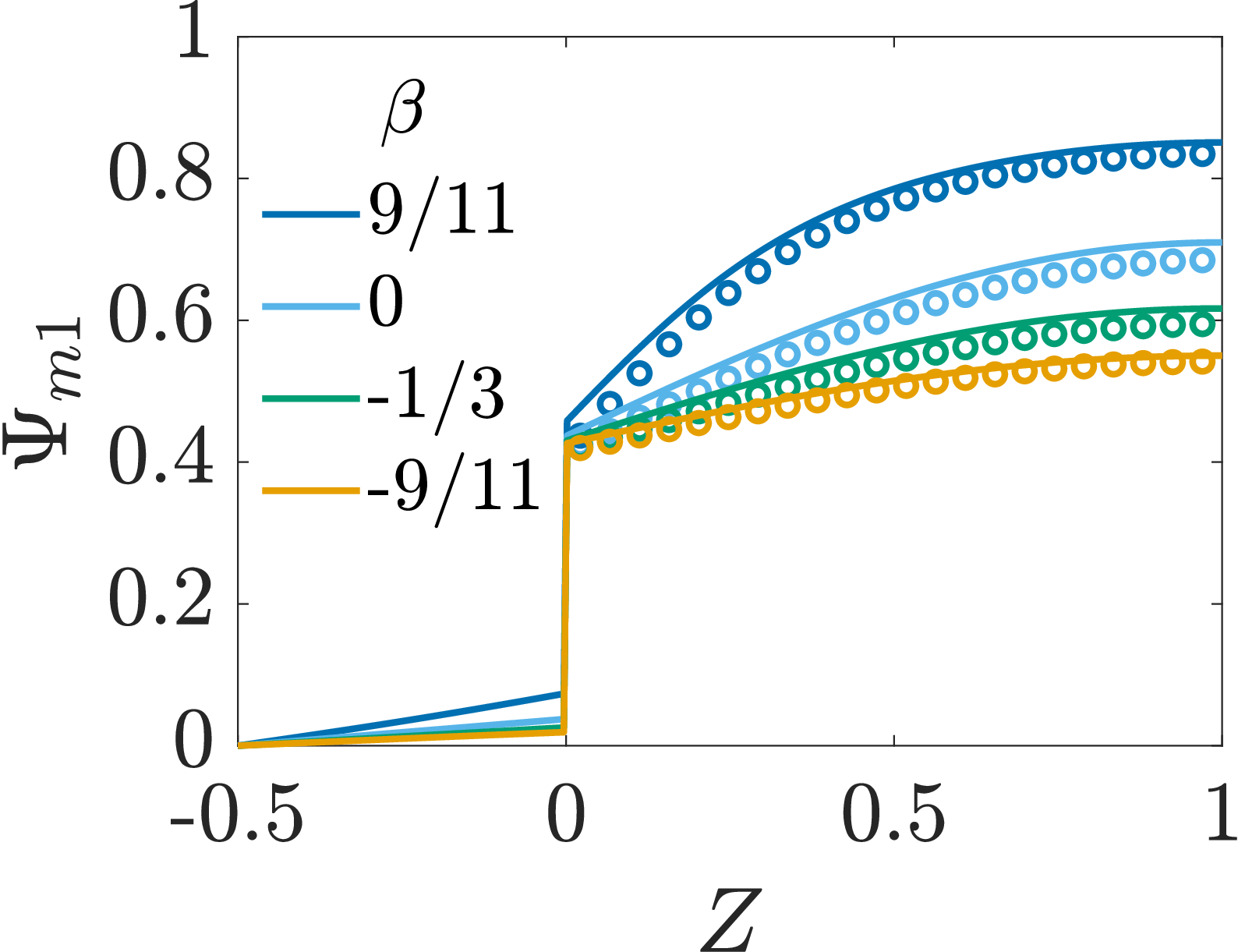}
    \end{subfigure}%
    \begin{subfigure}{0.33\textwidth}
    \centering
    \caption{}
    %\large{$\gamma=0$}\par\medskip
    \includegraphics[max width=.95\textwidth]{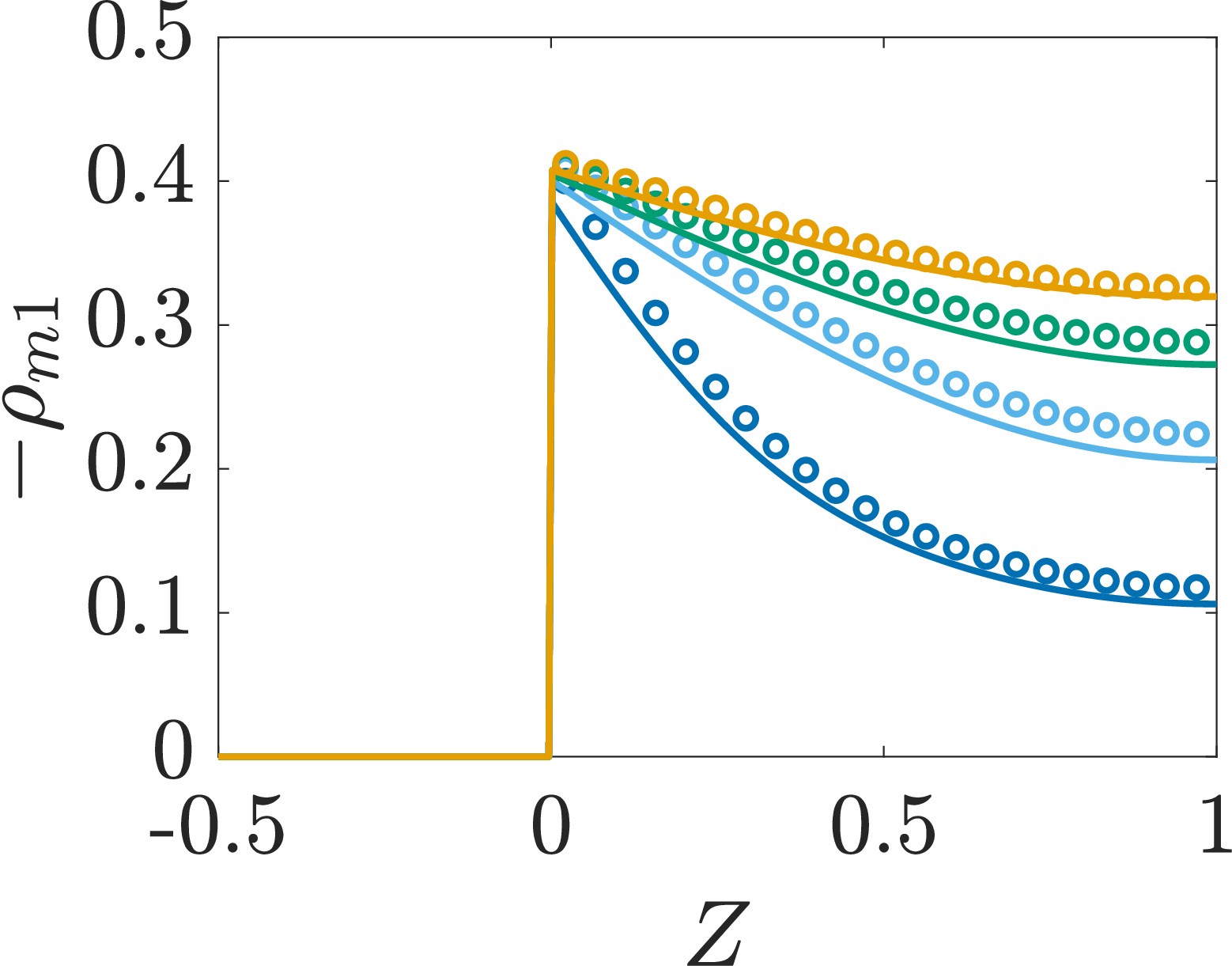}
    \end{subfigure}
    \begin{subfigure}{0.33\textwidth}
    \caption{}
    %\phantom{\large{$\gamma=0$}}\par\medskip
    \centering
    \includegraphics[max width=.95\textwidth]{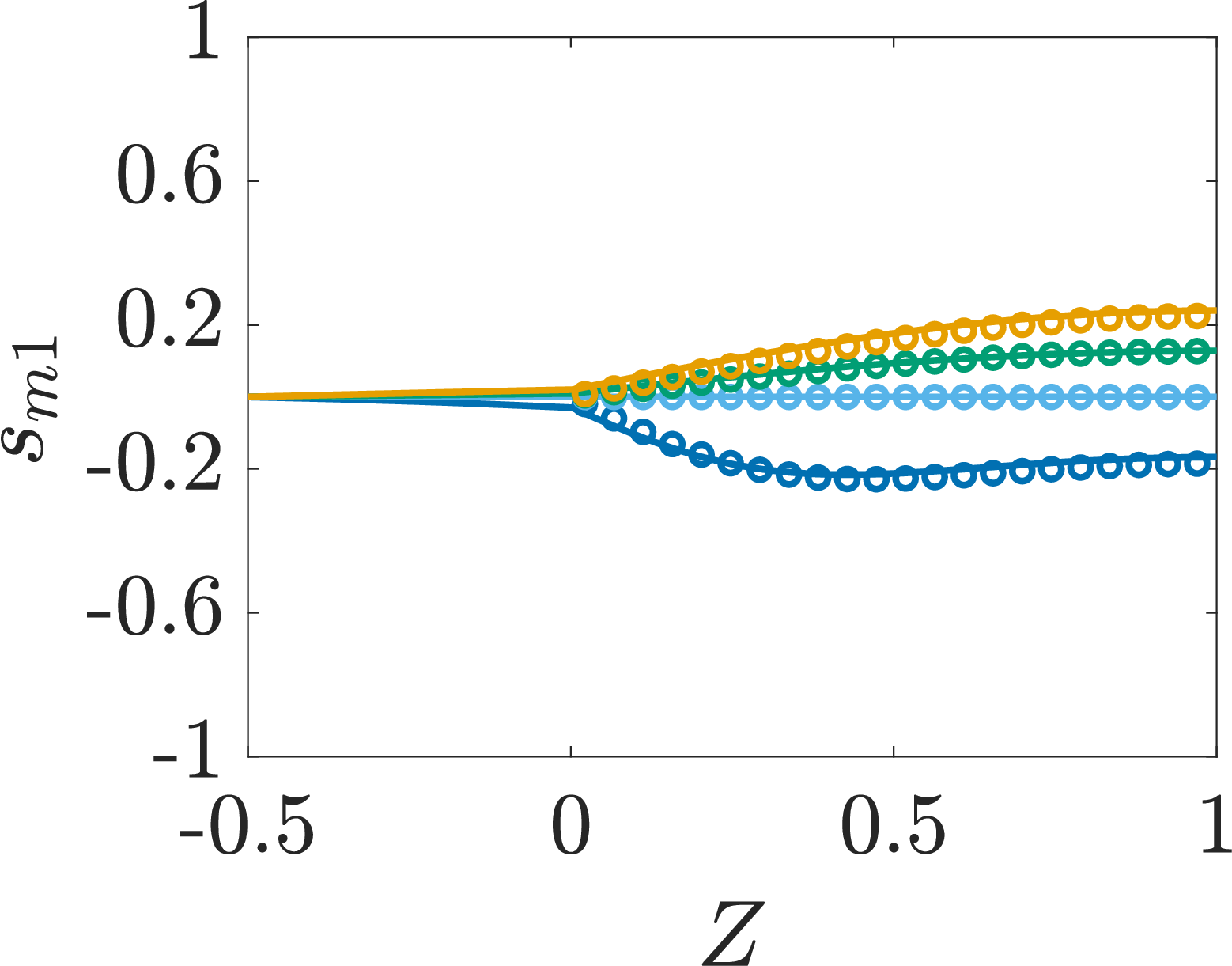}
    \end{subfigure}
    \caption{\textbf{Comparison between finite and infinite SDL conductance solutions of the perturbation model for different ion diffusivities}. Centerline profiles of the independent variables at $tD_+/\ell_p^2=0.28$ for different values of anion diffusivity: $D_-/D_+=0.1$ ($\beta=9/11$), $D_-/D_+=1$ ($\beta=0$), $D_-/D_+=2$ ($\beta=-1/3$) and $D_-/D_+=10$ ($\beta=-9/11$). Finite-difference solutions of the reduced-order model of Eqs. \eqref{eq:pnpsdlav} and \eqref{eq:pnpav} for $\alpha=10$ are represented by solid lines, while its analytical solutions for $\alpha\to\infty$ (Eqs. \eqref{Eq: pnp} and \eqref{Eq: fluxes}) are represented by circles. a) potential, b) charge density, c) salt density. $\gamma=0$ for all plots. The reduced-order model shows good agreement with the DNS. A contrast in diffusivities accelerates or hinders charging, in addition to producing appreciable salt enrichment or depletion in the SDL and pore regions.}
    \label{fig:profiles_highalpha}
\end{figure*}

Fig. \ref{fig:profiles_highalpha} shows a comparison of the analytical solutions for centerline charge, potential, and salt profiles in the limit of high SDL conductance  (Eqs.  \eqref{eq:rhobar1nosdl} and \eqref{eq:sbar1nosdl}, using Eqs. \eqref{eq:rho_bar_m}--\eqref{eq:s_bar_m}) with finite-difference solutions of the reduced-order model for $\alpha=10$ (Eqs. \eqref{eq:pnpav} and \eqref{eq:bulksalt}--\eqref{eq:psi0}). The profiles for this moderate value of SDL conductance agree well with the analytical solutions from the infinitely-conducting approximation. Comparing Fig. \ref{fig:profiles_highalpha}a to Fig. \ref{fig:profilesbetaZ}a, we see that the potential drops over the SDL are reduced as $\alpha$ is increased, rendering the approximation of trivial SDL dynamics more applicable. 

The results \eqref{eq:rhobar1nosdl} and \eqref{eq:sbar1nosdl} enable us to determine expressions for the charge and salt average fluxes, whose negative derivative figures on the right-hand side of Eq. \eqref{eq:highalpha}. These fluxes read
\begin{equation}
    \begin{bmatrix}
    \bar{J}_{1Z} \\[5pt]
    \bar{W}_{1Z}
    \end{bmatrix}=-A\dfrac{\partial }{\partial Z}\begin{bmatrix}
    \bar{\rho}_{1} \\[5pt]
    \bar{s}_{1}
    \end{bmatrix}.
    \label{eq:fluxes_nosdl}
\end{equation}
Before determining expressions for the fluxes, we note some physical consequences of Eq. \eqref{eq:fluxes_nosdl}. This result expresses the average charge and salt fluxes of asymmetric electrolytes as produced by linear combinations of both the average charge and salt density gradients. This leads to a natural classification of these fluxes as charge-induced or salt-induced, according to which gradient produces them.
Alternatively, due to the the possibility of combining potential and charge derivatives in the governing equations that arises from the radial equilibrium relations \eqref{eq:Requil} and \eqref{eq:poisson_sol} -- allowing the passage from Eq. \eqref{eq:pnpZ} to \eqref{eq:pnpav} -- we can relate the classification of the fluxes in charge- and salt-induced parts to the more commonly used representation in terms of diffusive and electromigrative fluxes,
\begin{equation}
    \bar{J}_{1Z}=-\underbrace{\left[(1+\beta\gamma)\dfrac{\partial\bar{\rho}_1}{\partial Z}+\beta(1-\gamma^2)\dfrac{\partial\bar{s}_1}{\partial Z}\right]\tau_0}_{\textrm{diffusion}}-\underbrace{\beta(1-\tau_0)\dfrac{\partial\bar{\rho}}{\partial Z}}_{\mathrm{electromigration}}=-\underbrace{(1+\beta\gamma)\dfrac{\partial\bar{\rho}_1}{\partial Z}}_{\mathrm{charge-induced}}-\underbrace{\beta(1-\gamma^2)\tau_0\dfrac{\partial\bar{s}_1}{\partial Z}}_{\mathrm{salt-induced}}
    \label{eq:Jdiffmig}
\end{equation}
and
\begin{equation}
    \bar{W}_{1Z}=-\underbrace{\left[\beta\dfrac{\partial\bar{\rho}_1}{\partial Z}+(1-\beta\gamma)\dfrac{\partial\bar{s}_1}{\partial Z}\right]\tau_0}_{\textrm{diffusion}}-\underbrace{\beta(1-\tau_0)\dfrac{\partial\bar{\rho}}{\partial Z}}_{\mathrm{electromigration}}=-\underbrace{\beta\dfrac{\partial\bar{\rho}_1}{\partial Z}}_{\textrm{charge-induced}}-\underbrace{(1-\beta\gamma)\tau_0\dfrac{\partial\bar{s}_1}{\partial Z}}_{\textrm{salt-induced}}.
    \label{eq:Wdiffmig}
\end{equation} 
Eqs. \eqref{eq:Jdiffmig} and \eqref{eq:Wdiffmig} support the argument that diffusion dominates for narrow pores (when $\kappa\ll 1$ and $\tau_0\sim 1$) and migration for wide pores ($\kappa\gg 1$ and $\tau_0\ll 1$). Bearing in mind the linear relation between the derivatives of average charge and average potential, we can interpret charge-induced fluxes as effective electromigrative fluxes, and salt-induced fluxes as related to concentration overpotentials. Thus, even in the overlapping double layer limit $\kappa\ll 1$, charge-induced fluxes are at play acting as effective electromigration fluxes.

\begin{figure}[t!]
    \centering
    \includegraphics[max width=\textwidth]{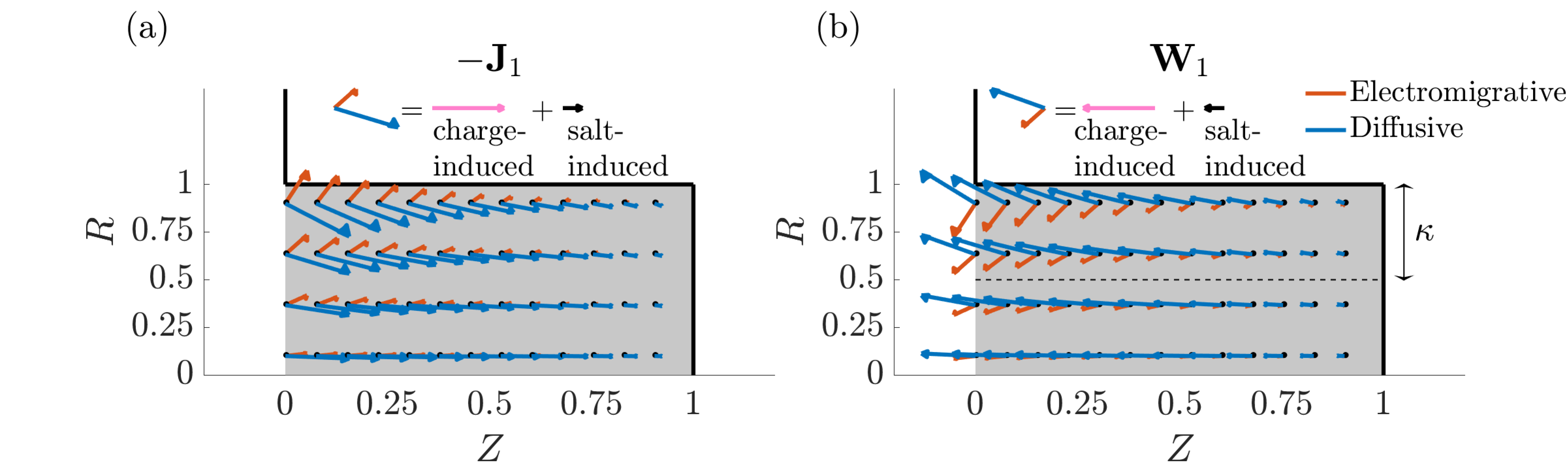}
    \caption{\textbf{Diffusive and electromigration fluxes in the pore}. Vector field plots of diffusive and electromigrative fluxes and an illustration of their regrouping into charge- and salt-induced fluxes. a) Negative diffusive and electromigrative charge fluxes in the pore, $-\mathbf{J}_1$, b) positive diffusive and electromigrative salt fluxes in the pore, $\mathbf{W}_1$.
    Arrow lengths are linearly scaled. $\beta=2/3$, $\gamma=0$, $T=0.1$, and $\kappa=2$ for both plots. The balance of radial fluxes sets equilibrium in that direction, while axial fluxes, constant across the cross section, give rise to charge and salt dynamics. Charge- and salt-induced fluxes are purely axial, and the latter may compete or cooperate with charging.}
    \label{fig:flux_quiver}
\end{figure}

Now, plugging Eqs. (\ref{eq:rhobar1nosdl}) and (\ref{eq:sbar1nosdl}) into \eqref{eq:fluxes_nosdl}, we differentiate term-wise to obtain
\begin{equation}
\begin{bmatrix}
    \bar{J}_{1Z} \\[5pt]
    \bar{W}_{1Z}
    \end{bmatrix}=2A\tau_0\sum_{n=1}^\infty\cos(\xi_n Z)\begin{bmatrix}
    g_n(\mathcal{T}) \\[5pt]
    h_n(\mathcal{T})
    \end{bmatrix}.
    \label{eq:fluxes_nosdl_soln}
\end{equation}
 Fig. \ref{fig:flux_quiver} shows the diffusive and electromigrative fluxes constructed from the axial components\footnote{Note that Eq. \eqref{eq:Requil} implies that these fluxes are constant over the cross section, and equal to their average values.} in Eqs. \eqref{eq:fluxes_nosdl_soln}, and the radial components in Eq. \eqref{eq:rad_equil}, calculated using \eqref{eq:Requil} and \eqref{eq:poisson_sol}, for an electrolyte with asymmetric diffusivities ($\beta=2/3$) and a moderate relative pore size ($\kappa=2$). The figure shows that the radial components of the diffusive and electromigrative charge and salt fluxes are in balance at all positions in the pore, promoting radial equilibrium across the small transversal direction. As for symmetric electrolytes \cite{henrique2022charging}, the axial components of the diffusive and electromigrative charge fluxes cooperate to set the rate of axial charge transport. However, in contrast to that scenario, a mismatch in diffusivities produces electromigration of salt, which in turn induces a previously unobserved salt dynamics. The representation of charge and salt fluxes in terms of  charge- and salt-induced parts is also shown in the figure. An important feature of this subdivision is the absence of radial components in each of those parts. Since all radial fluxes are accounted for as charge-induced contributions, they sum up to zero. From this perspective, the novel feature of the model for asymmetric electrolytes is the presence of salt-induced fluxes.

\begin{figure}[t!]
    \centering
    \includegraphics[max width=.95\textwidth]{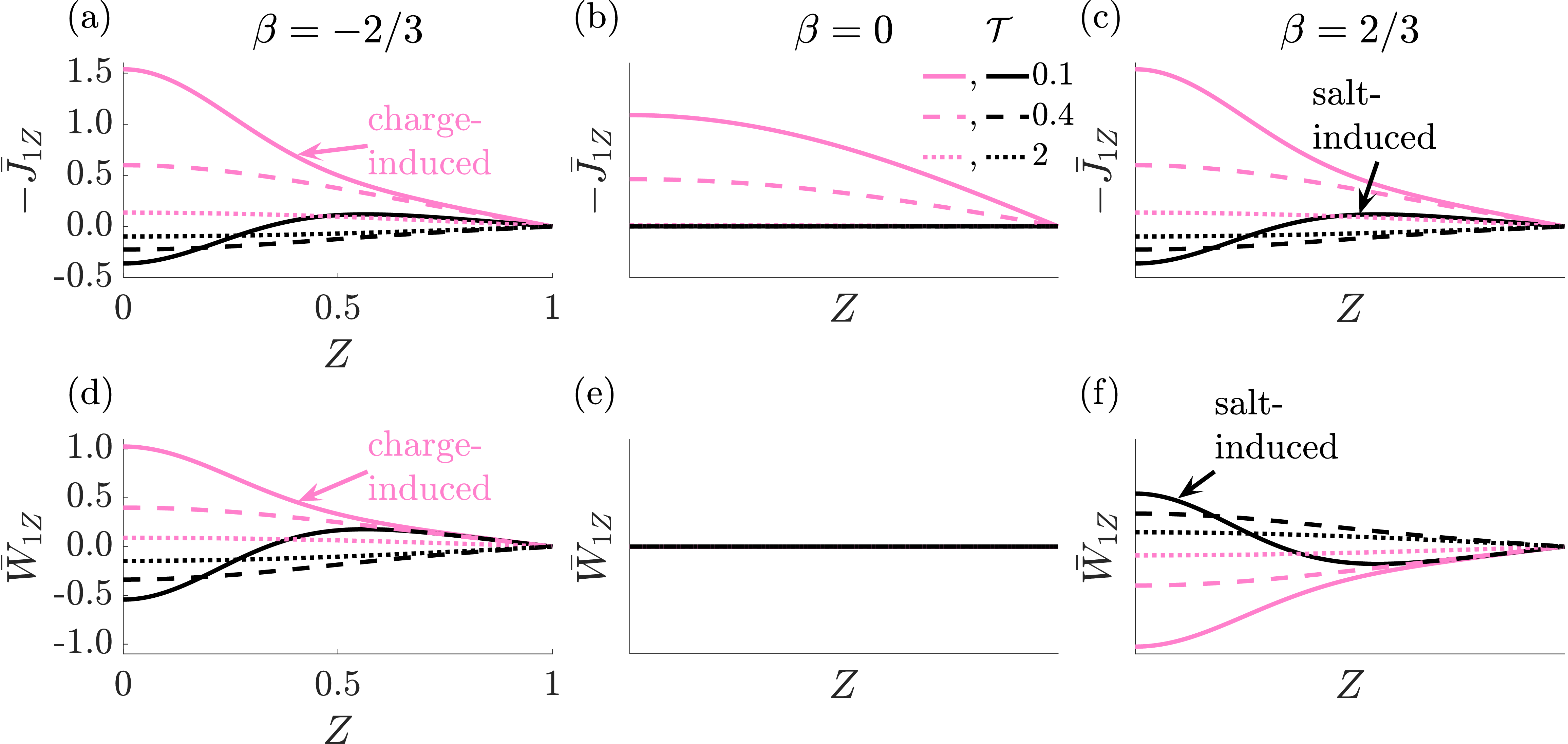}
    \caption{\textbf{Profiles of charge- and salt-induced fluxes in the pore for different ion diffusivities at distinct times}. Charge- (in pink) and salt-induced (in black) fluxes of charge and salt across the pore for different times and diffusivities mismatch parameters. a)--c) charge flux components for $\beta=-2/3$, $\beta=0$ and $\beta=2/3$, respectively. d)--f) salt flux components for $\beta=-2/3$, $\beta=0$ and $\beta=2/3$, respectively. $\kappa=2$ and $\gamma=0$ for all plots, which share the same $Z$ axis.}
    \label{fig:fluxes_asym}
\end{figure}

To shed light onto the effects of charge- and salt-induced fluxes on the ion transport dynamics, we plot both of these contributions to charge and salt fluxes across the pore in Fig. \ref{fig:fluxes_asym}. We note that as previously stated in the end of Sec. \ref{sec:form}, the symmetry of the system with respect to the inversion of the roles of cations and anions is such that when the sign of $\beta$ is reversed (since in this case, for $\gamma=0$), charge flux remains the same, but the sign of salt flux flips. Fig. \ref{fig:fluxes_asym} illustrates the monotonicity of all charge-induced fluxes, continuously introducing charge into the pore (Figs. \ref{fig:fluxes_asym}a--c) and enriching salt in the pore when anions outpace cations (Fig. \ref{fig:fluxes_asym}d) or depleting it in the opposite case (Fig. \ref{fig:fluxes_asym}f). As previously reported \cite{gupta2020charging,henrique2022charging}, in the symmetric case (Fig. \ref{fig:fluxes_asym}e), no salt dynamics occurs. Salt-induced fluxes, only present for asymmetric electrolytes, present a more interesting behavior. At short times they are non-monotonic, competing with charge-induced fluxes near the mouth of the pore, promoting discharging (Figs. \ref{fig:fluxes_asym}a,c) and hindering salt transport produced by charge-induced fluxes for either enrichment or depletion (Figs. \ref{fig:fluxes_asym}d,f).  Closer to the end of the pore, they cooperate with charge-induced fluxes, favoring charging and collaborating with the direction of salt transport induced by the latter. At long times, however, salt-induced fluxes become monotonic and compete with both the charge and salt charge-induced fluxes throughout the entire pore, slowing down the late dynamics. It should be noted that the early dynamics of the system is sped up since the gains in charge-induced fluxes outweigh the losses due to salt-induced ones. For brevity, we keep the discussion on the mechanistic implications of simultaneous asymmetries in valences and diffusivities to \ref{Sec:app}. 

A natural question arises: under what conditions can the net effect of electrolyte asymmetry accelerate the charging dynamics? To address it, we note that Eqs. (\ref{eq:rhobar1nosdl}) and (\ref{eq:sbar1nosdl}) furnish a straightforward way of deriving the effect of parameter dependence on net properties, such as a characteristic time required to approach steady-state charge storage, and the maximum salt variation in the pore, in this regime of negligible SDL resistance. Those properties are determined from the total dimensionless charge $Q_1=\int_0^1\bar{\rho}_1\,dZ$ and salt $S_1=\int_0^1\bar{s}_1\,dZ$ stored in the pore, yielding
\begin{equation}
    Q_1(\mathcal{T})=\tau_0\left(-1+2\sum_{n=1}^\infty \dfrac{g_n(\mathcal{T})}{\xi^2_n}\right)
\end{equation}
and
\begin{equation}
    S_1(\mathcal{T})=2\tau_0\sum_{n=1}^\infty\dfrac{h_n(\mathcal{T})}{\xi^2_n}.
\end{equation}
\begin{figure}[t!]
    \centering
    \includegraphics[max width=.95\textwidth]{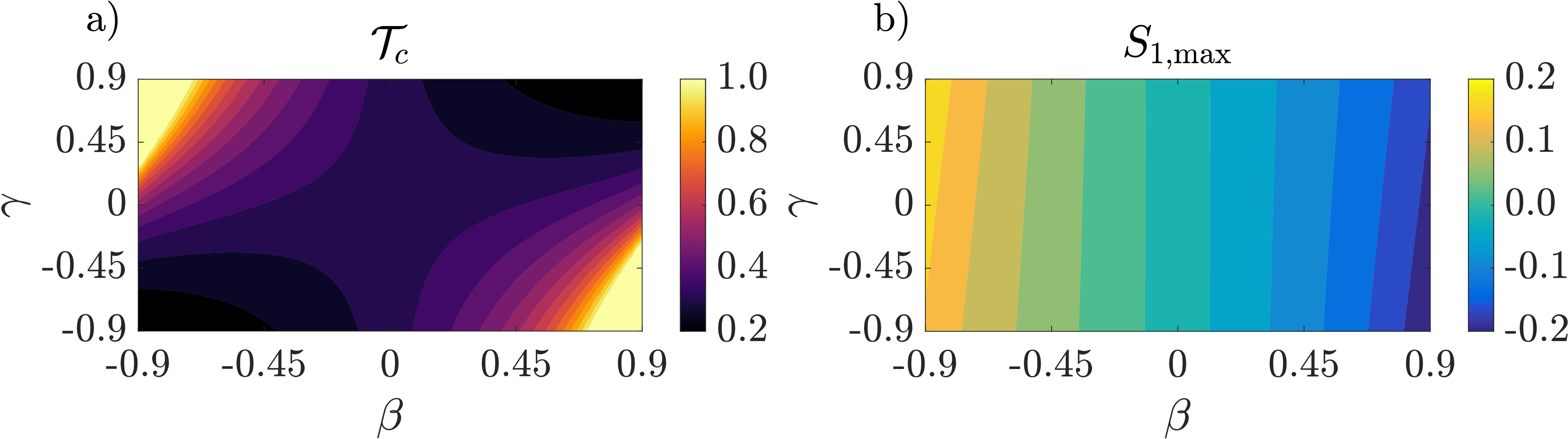}
    \caption{\textbf{Net charging properties}. Contour plots of the net charging timescale $\mathcal{T}_c$ and maximum salt change in the pore $S_{1,\textrm{max}}$ for different contrasts in valences $\gamma$ and diffusivities $\beta$. Charging is limited by the slowest timescale in Eq. (\ref{eq:eigA}), and can be optimized by concurrent increases in both the valence and diffusivity of either ionic species.}
    \label{fig:taucs}
\end{figure}
From these results, we determine 
a charging timescale of the pore $\mathcal{T}_c$ and the maximum total salt extracted from or inserted into the pore, $S_{1,\mathrm{max}}$. We define $\mathcal{T}_c$ as the dimensionless time taken for the pore to attain a ratio of $1-e^{-1}$ of its steady-state charge. The dependences of both properties on diffusivity and valence asymmetries are shown in Fig. \ref{fig:taucs}. These results indicate that the short-time charge-induced flux boost accelerates charging for high like-signed contrasts. However, the higher salt-induced fluxes occur for high oppositely-signed contrasts, as rationalized in the discussion of Fig. \ref{fig:fluxes_asym} and further discussed in \ref{Sec:app}. Thus, an interesting takeaway from Fig. \ref{fig:taucs} is the route to optimize the charging timescale of pores: asymmetry of diffusivities alone introduces losses due to concentration overpotentials that slow down the overall charging. However, a suitable combination of commensurate like-signed increases in diffusivity and valence contrasts leads to enhanced electromigrative mobility, and thus accelerated charging.

\section{Conclusion}

In this work, we extended the approach of Refs. \cite{gupta2020charging,henrique2022charging} to describe effects of valence and diffusivity asymmetry (i.e., electrolyte asymmetry) on the charging of binary dilute electrolytes in the Debye-H{\"u}ckel limit. Importantly, we noted that in this case, the mismatch in ionic diffusivities produces salt migration. Through an averaging of the Poisson-Nernst-Planck equations, we reach coupled transport equations for average charge and salt densities, and electric potential. This set of equations is solved by the finite-difference method at a low cost and we find good agreement between its results and the direct numerical simulations. 

A simplified version of model is obtained when the resistance of the static diffusion layer may be neglected, allowing for an analytical solution of the governing equations. We show that this approximation is in quantitative agreement with numerical solutions for $\alpha=10$. From the analytic solution, we derive the dependence of charge flux on the mismatch of valences and asymmetries. 
A key takeaway from our analysis is the existence of two distinct charging timescales for asymmetric electrolytes in cylindrical pores. With our solution, we quantify this dependence on relative pore size, and show that in the limit of overlapping double layers, it reduces to the diffusion timescales of cations and anions, $\ell_p/D_\pm$. For intermediate pore sizes, these timescales assume a more intricate dependence that will be further explored in future work.

Physically, we discuss how the electromigrative and diffusive fluxes may be decomposed into charge-induced, resulting in an effective electromigrative contribution, and salt-induced, related to a concentration overpotential. We noted that 
charge-induced fluxes dominate at short times, controlling the rate of charge flux, and inducing either salt enrichment or depletion in the pore. An important conclusion is the possibility of accelerating charging through a simultaneous increase in the diffusivity and valence of either species. Remarkably, while the capacitance of the pore is independent of transport coefficient asymmetries, the charging rate of the pore is not constrained to the steady-state charge storage.

It should be noted that formally, as discussed by Aslyamov and Janssen \cite{aslyamov2022analytical}, the method discussed on this paper and on Refs. \cite{gupta2020charging,henrique2022charging} is based on an additional asymptotic expansion for pores with large aspect ratios, much longer than wide, and can be extended to the nonlinear potential regime by obtaining exact solutions of the Poisson-Boltzmann equation for the electric potential along the rapidly equilibrating cross-section of the long pore.

While the results described here focus on a binary electrolyte, this framework can be extended to tackle a number of applications to electrochemical systems. To this end, we can reformulate it for an arbitrary number of ions to address multicomponent electrolytes \cite{zhao2012time,jarvey:submitted}, incorporate surface reduction-oxidation reactions \cite{biesheuvel2011diffuse,jarvey:submitted} to model hybrid capacitors, and account for a time-dependent applied potential to predict impedance spectra \cite{song2012effects,mei2018physical,tomlin2021impedance,balu2022electrochemical} and cyclic voltammetry \cite{wang2012physical,yan2017theory,mei2017three} curves of electrochemical devices.  In summary, our work provides fundamental understanding of  the impact asymmetry in cation and anion diffusivities on the behavior of ionic transport inside a charged cylindrical pore and paves the way for future research on electrolyte transport in a variety of electrochemical systems that consist of porous materials.  

\section*{Conflicts of Interest}
There are no conflicts to declare.

\section*{Acknowledgements}
The authors would like to acknowledge the helpful input provided by Gesse Roure, Howard Stone, and Robert Davis. The authors acknowledge that the simulations reported in this contribution were performed using the
Princeton Research Computing resources at Princeton University which is consortium of groups including the Princeton Institute for Computational Science and Engineering and the Princeton University Office of Information Technology’s Research Computing department. F. H. and A. G. would like to thank the Ryland Family Graduate Fellowship for financial support.

\appendix

\setcounter{figure}{0}
\renewcommand\thefigure{A\arabic{figure}}

\section{Fluxes with Valence and Diffusivity Asymmetries in Tandem}\label{Sec:app}

\begin{figure}[t!]
    \centering
    \includegraphics[max width=\textwidth]{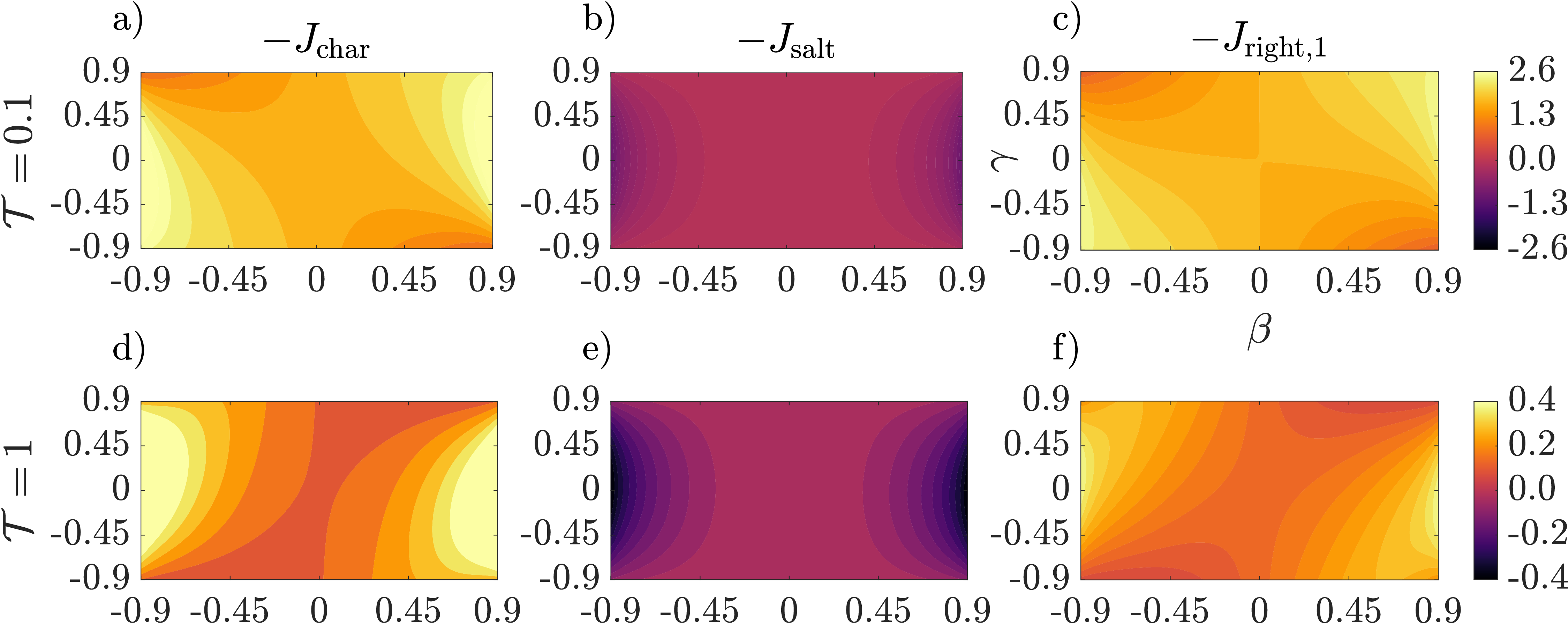}
    \caption{\textbf{Dependence of charge flux at the mouth of the pore on asymmetries in valences and diffusivities}. Contour plot of charge- and salt-induced charge fluxes for negligible SDL resistance. $\kappa=2$. Like-signed increases in diffusivity and valence asymmetries (top right and bottom left) boost charge-induced flux and speed up charging. In the long-time dynamics, charge- and salt-induced fluxes compete to set the common steady-state charge profile.}
    \label{fig:charctr}
\end{figure}

\begin{figure}[t!]
    \centering
    \includegraphics[max width=\textwidth]{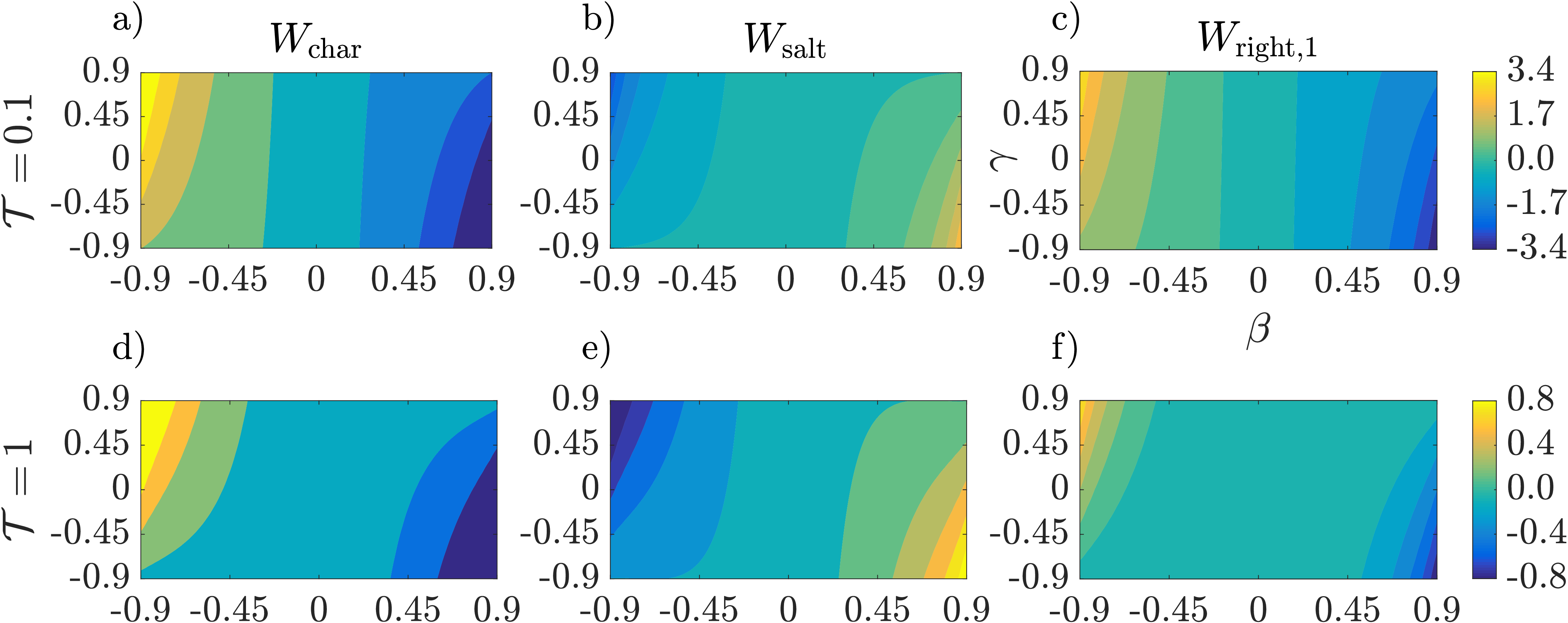}
    \caption{\textbf{Dependence of salt flux at the mouth of the pore on asymmetries in valences and diffusivities}. Contour plot of charge- and salt-induced salt fluxes for negligible SDL resistance. $\kappa=2$. Like-signed increases in diffusivity and valence asymmetries (top right and bottom left) screen the potential faster. Therefore, while increases in diffusivity contrast enhance salt flux, like-signed increases in valence contrast reduce the driving force for salt transport. Thus, salt transport is boosted for for low $\beta\gamma$ (top left and bottom right). In the long-time dynamics, charge- and salt-induced fluxes compete to set the common steady-state charge profile.}
    \label{fig:saltctr}
\end{figure}

Let us denote charge- and salt-induced fluxes respectively by the indices ``char'' and ``salt''. Such contributions to the total charge and salt fluxes at the mouth of the pore are shown in Figs. \ref{fig:charctr} and \ref{fig:saltctr}, respectively. In each of these figures, the first row shows a prediction of the domination of charge-induced fluxes at short times. This occurs due to the initial uniformity of both ionic concentration profiles, yielding small salt inhomogeneities. Due to the rapid radial equilibration for long pores, the applied potential quickly induces a combination of electromigrative and charge-induced diffusive fluxes which dominates salt-induced diffusion at that timescale. 

Fig. \ref{fig:charctr}a shows higher charge-induced charge flux on the top right and bottom left corners. This indicates that increases in both the diffusivity and valence of either ionic species, or equivalently in the product $\beta\gamma$, boost the dominating electromigrative charge transport in this short-time regime. On the other hand, Fig. \ref{fig:charctr}b illustrates the weak dependence of salt-induced charge flux on valence asymmetry for short times. However, the increase in charge flux for higher $\beta\gamma$ also wears off faster.

% In fact, the charge-induced salt transport coefficient $\beta$ is not a function of valence asymmetry, and such valence dependences arise over time due to coupling of salt and potential.

The quicker potential screening observed for higher $\beta\gamma$ produces a decrease of charge-induced salt flux when $\beta$ is kept constant and $\mathrm{sgn}(\beta)\gamma$ increases. That is to say that since the charge gradient profile flattens out more quickly in the top right and bottom left of Fig. \ref{fig:charctr}a, the driving force for salt transport is reduced in those regions of the parameter space, so salt transport is higher in the opposite corners, top left and bottom right of Fig. \ref{fig:saltctr}c. Salt enrichment occurs when anion electromigrative attraction is faster than cation repulsion, i.e., for $\beta>0$, and depletion when $\beta<0$; see Eq. (\ref{eq:pnpav}).

The second rows of Figs. \ref{fig:charctr} and \ref{fig:saltctr} show that the charge and salt contributions to both charge and salt fluxes become comparable at long times, as a result of the salt inhomogeneity that has been promoted by the cooperating charge-induced diffusive and electromigrative fluxes across the entire asymmetry parameter space. As shown in Fig. \ref{fig:charctr}e, salt-induced charge flux hinders charging, smoothing out the charge density profile. Fig. \ref{fig:saltctr}e displays a similar trend, where salt-induced salt fluxes approach charge-induced ones. In this case, the former overcome the latter to promote a uniform salt profile. Indeed, at long times, all charge- and salt-induced fluxes decay at comparable magnitudes until the common steady-states $\bar{\rho}_{1}(Z,\mathcal{T}\to\infty;\beta,\gamma)=-\tau_0$ and $\bar{s}_{1}(Z,\mathcal{T}\to\infty;\beta,\gamma)=0$ (see Eqs. (\ref{eq:rhobar1nosdl}) and (\ref{eq:sbar1nosdl})) are reached for all values of diffusivity and valence contrast.

Figs. \ref{fig:charctr} and \ref{fig:saltctr} provide the means to appreciate symmetry in parameter dependence and its implications. We remark the symmetries $\bar{J}_{1Z}(Z,\mathcal{T};-\beta,-\gamma)=\bar{J}_{1Z}(Z,\mathcal{T};\beta,\gamma)$ and $\bar{W}_{1Z}(Z,\mathcal{T};-\beta,-\gamma)=-\bar{W}_{1Z}(Z,\mathcal{T};\beta,\gamma)$, mentioned at the end of Sec. \ref{sec:IC}. Physically, they correspond to an inversion of the roles of cations and anions: if both their valences and diffusivities were switched, cations would be repelled from the pore at the rate that anions are attracted, and vice versa. This would produce the same electric potential, but the opposite salt change, resulting in the relations $\bar{\Psi}_{1}(\mathcal{T};\beta,\gamma)=\bar{\Psi}_{1}(\mathcal{T};-\beta,-\gamma)$, $\bar{s}_{1}(\mathcal{T};\beta,\gamma)=-\bar{s}_{1}(\mathcal{T};-\beta,-\gamma)$.

% As time increases, potential is screened, such that the charge and salt profiles in the pore saturate and the maximum charge and salt fluxes change. 

\pagebreak

\bibliography{apssamp}% Produces the bibliography via BibTeX.

% \appendix

% \setcounter{figure}{0}
% \renewcommand\thefigure{A\arabic{figure}}

% \section{Transmission Line Circuit Resistances}

% \section{Areal Capacitance}

% \section{Electrode Average Properties} 

\end{document}